\documentclass[12pt]{iopart}
\usepackage{txfonts,color}
\usepackage{graphicx}
\usepackage{iopams}
\usepackage{setstack}
\usepackage{stmaryrd}
\usepackage{multirow,color}

\begin{document}

\title{Order flow dynamics around extreme price changes on an emerging stock market}

\author{Guo-Hua Mu $^{1,2,3}$, Wei-Xing Zhou $^{1,2,4,5}$, Wei Chen $^{6}$ and J{\'a}nos Kert{\'e}sz $^{3,7}$}

\address{$^1$ School of Business, East China University of Science and Technology, Shanghai 200237, People's Republic of China}
\address{$^2$ School of Science, East China University of Science and Technology, Shanghai 200237, People's Republic of China}
\address{$^3$ Department of Theoretical Physics, Budapest University of Technology and Economics, Budapest, H-1111, Hungary} %
\address{$^4$ Research Center for Econophysics, East China University of Science and Technology, Shanghai 200237, People's Republic of China}
\address{$^5$ Research Center on Fictitious Economics \& Data Science, Chinese Academy of Sciences, Beijing 100080, People's Republic of China}%
\address{$^6$ Shenzhen Stock Exchange, Shenzhen 518010, People's Republic of China} %
\address{$^7$ HAS-BME Condensed Matter Research Group, Budapest, H-1111, Hungary} %

\ead{wxzhou@ecust.edu.cn (W.-X. Zhou), kertesz@phy.bme.hu (J. Kert{\'e}sz)}

\begin{abstract}
We study the dynamics of order flows around large intraday price changes using ultra-high-frequency data from the Shenzhen Stock Exchange. We find a significant reversal of price for both intraday price decreases and increases with a permanent price impact. The volatility, the volume of different types of orders, the bid-ask spread, and the volume imbalance increase before the extreme events and decay slowly as a power law, which forms a well-established peak. The volume of buy market orders increases faster and the corresponding peak appears earlier than for sell market orders around positive events, while the volume peak of sell market orders leads buy market orders in the magnitude and time around negative events. When orders are divided into four groups according to their aggressiveness, we find that the behaviors of order volume and order number are similar, except for buy limit orders and canceled orders that the peak of order number postpones two minutes later after the peak of order volume, implying that investors placing large orders are more informed and play a central role in large price fluctuations. We also study the relative rates of different types of orders and find differences in the dynamics of relative rates between buy orders and sell orders and between individual investors and institutional investors. There is evidence showing that institutions behave very differently from individuals and that they have more aggressive strategies. Combing these findings, we conclude that institutional investors are more informed and play a more influential role in driving large price fluctuations.
\end{abstract}

\submitto{\NJP}

\maketitle

{\color{blue}{\tableofcontents}}

\section{Introduction}
\label{S1:Intro}

The current financial tsunami, which is regarded as the biggest one after the Great Depression, has attracted much interest not only in the industries but also in the academic community. The direct trigger of the financial breakdown was the subprime mortgage crisis in the United States of America, and there are numerous efforts aiming to explain its causes and paths. From a complex systems point of view, the financial tsunami has nothing special since ``{\em{the accumulation of several bubbles and their interplay and mutual reinforcement has led to an illusion of a `perpetual money machine' allowing financial institutions to extract wealth from an unsustainable artificial process}}'' \cite{Sornette-Woordard-2009-XXX}. In this framework, unfolding the crisis is related at least to the speculative bubble of foreign capital inflow to the U.S.A. \cite{Sornette-Zhou-2004-PA}, the ``new economy'' ICT bubble \cite{Johansen-Sornette-2000a-EPJB}, the slaving of the U.S.A Federal Reserve to the stock market \cite{Zhou-Sornette-2004b-PA}, the real-estate bubble \cite{Zhou-Sornette-2006b-PA}, and the commodity bubbles \cite{Sornette-Woordard-Zhou-2009-PA}. These phenomena are identified as bubbles and crashes \cite{Sornette-2003-PR}.

More generally, one of the focal issues of economics is the understanding of the price formation mechanism. By now much is known about the stylized facts related to the price fluctuations (see, e.g., \cite{Cont-2001-QF,Johnson-Jefferies-Hui-2003}), but we are far from having a clear picture about the origins. As illustrated by the events mentioned in the previous paragraph, this is a question of general public interest. It is natural to assume, especially from the physics point of view, that the behavior around major changes will tell us important related details. As we would like to take a statistical approach, we have to restrict ourselves to ``large changes on small scales'', i.e., to the question: What happens in the neighborhood of a {\it relatively} large change if high resolution data are considered.

An early work about aftershock market behavior is due to Lillo and Mantegna who studied the relaxation dynamics of the occurrence of large volatility after volatility shocks, as a analogue of the Omori law after earthquakes \cite{Lillo-Mantegna-2004-PA}. They investigated 1-minute logarithmic changes of the S\&P 500 index during 100 trading days after the Black Monday and found that the occurrence of events larger than some threshold exhibits power-law relaxation for different thresholds \cite{Lillo-Mantegna-2004-PA}. Sel\c{c}uk investigated daily index data from 10 emerging stock markets (Argentina, Brazil, Hong Kong, Indonesia, Korea, Mexico, Philippines, Singapore, Taiwan and Turkey) and observed Omori's law after two largest crashes in each market \cite{Selcuk-2004-PA}. Sel\c{c}uk and Gen\c{c}ay utilized the 5-minute Dow Jones Industrial Average 30 index (DJIA) and identified the Omori law after October 8, 1998 and January 3, 2001 \cite{Selcuk-Gencay-2006-PA}. In addition, the power-law relaxation could happen after intermediate shocks, which was confirmed in the USA markets \cite{Weber-Wang-VodenskaChitkushev-Havlin-Stanley-2007-PRE} and in the Chinese markets \cite{Mu-Zhou-2008-PA,Jiang-Li-Cai-Wang-2009-PA}.

Another related topic is to study the relaxation dynamics of financial measures after large price changes. Sornette and coworkers found that the implied variance of the Standard and Poor's 500 (S\&P 500) index after the infamous Black Monday (19 October 1987) decays as a power law decorated with log-periodic oscillations \cite{Sornette-Johansen-Bouchaud-1996-JPIF}.  Zawadowski et al examined the evolution of price, volatility and the bid-ask spread after extreme 15 min intraday price changes on the New York Stock Exchange (NYSE) and the NASDAQ \cite{Zawadowski-Kertesz-Andor-2004-PA}. They found a well-established price reversal and the volatility which increases sharply at the event decays according to a power law with an exponent of about 0.4. Zawadowski et al further showed that the volume and, in the case of the NYSE the bid-ask spread, which increase sharply at the event, stay significantly high over days afterwards and the decay of the volatility follows a power law \cite{Zawadowski-Andor-Kertesz-2006-QF}. T{\'o}th et al found similar power-law relaxations after large intraday price changes in the volatility, the bid-ask spread, the bid-ask imbalance, the number of queuing limit orders, the activity (number and volume) of limit orders placed and canceled in the stocks traded on the London Stock Exchange (LSE) \cite{Toth-Kertesz-Farmer-2009-EPJB}. Very recently, Ponzi et al. performed an analysis studying possible market strategies around large events \cite{Ponzi-Lillo-Mantegna-2009-PRE} and they found empirically that the bid-ask spread and the mid-price decay very slowly to their normal values when conditioned to a sudden variation of the spread. These findings show that the resiliency is slower than the assumed exponential form \cite{Obizhaeva-Wang-2008-JFinM,Alfonsi-Fruth-Schied-2010-QF}.

So far only mature markets have been studied with the technique of the statistical analysis of large changes in high resolution data. The aim of the present paper is to study an emergent market, namely the Chinese stock market from this point of view. We investigate the dynamics of several financial quantities around large price changes with special emphasis put on the order volume using a data base of the order book data of 23 stocks from the Shenzhen Stock Exchange (SZSE). More interestingly, the ultra-high-frequency data allow us to investigate the effect of order directions (buy and sell), order aggressiveness (partially filled orders, filled orders, limit orders and canceled orders as explained in Sec.~\ref{S1:Data}) and investor types (individual vs. institution) on the volume dynamics around extreme events and the behavior of relative rates of orders around large price changes.

The paper is organized as follows. Section \ref{S1:Data} presents a brief description of the data base we investigated. In Sec.~\ref{S1:DefineEvents}, we explain the definition of large prices changes and show that there is a price reversal after price jumps. Section \ref{S1:4measures} studies the dynamics of the absolute return, the volume, the bid-ask spread and the volume imbalance around large price changes. Sections \ref{S1:Buy:Sell}, \ref{S1:Agg}, and \ref{S1:Volume:Ind:Ins} investigate respectively the effect of order directions, order aggressiveness and investor types on the volume dynamics around extreme events and the behavior of relative rates of orders around large price changes. Section \ref{S1:conclusion} summarizes our findings.

\section{Data sets}
\label{S1:Data}

We use a database recording all orders of submission and cancelation of 23 liquid stocks traded on the SZSE in the whole year 2003. The market consists of three time periods on each trading day, namely, the opening call action (9:15 AM to 9:25 AM), the cooling period (9:25 AM to 9:30 AM), and the continuous double auction (9:30 AM to 11:30 AM and 1:00 PM to 3:00 PM). In this paper, we consider only the transactions occurring in the continuous double auction. The time, price and size of each order is recorded in the data set. In the Chinese stock market, the size of a buy order is limited to a board lot of 100 shares or an integer multiple thereof, while a seller can place a sell order with any size. The recorded trade size is in units of shares rather than board lots. The data set also contains the investor type for each order indicating whether the investor is an individual or an institution.

Note that only limit orders were allowed to submit in 2003. There are situations that a limit order is only partially executed since the total outstanding volume of orders waiting in the opposite order book with the prices no more aggressive than the submitted order is less than the size of the submitted order. Hence, only a part of shares are executed and the remaining shares are stored in the limit order book. This order is treated as two orders, one as a partially filled order and the other as a effective limit order. If the price of a buy (resp. sell) order is higher (resp. lower) than the best ask (resp. bid) price, the limit order is called an effective market order or a marketable order. The remaining limit orders are called effective limit orders. Without loss of clarity, we term these two types of orders as market orders and limit orders in this work. A market order is either a filled order or a partially filled order.

The tickers of the 23 stocks investigated are the following: 000001 (Shenzhen Development Bank Co. Ltd), 000002 (China Vanke Co. Ltd), 000009 (China Baoan Group Co. Ltd), 000012 (CSG holding Co. Ltd), 000016 (Konka Group Co. Ltd), 000021 (Shenzhen Kaifa Technology Co. Ltd), 000024 (China Merchants Property Development Co. Ltd), 000027 (Shenzhen Energy Investment Co. Ltd), 000063 (ZTE Corporation), 000066 (Great Wall Technology Co. Ltd), 000088 (Shenzhen Yan Tian Port Holdings Co. Ltd), 000089 (Shenzhen Airport Co. Ltd), 000406 (Sinopec Shengli Oil Field Dynamic Group Co. Ltd),000429 (Jiangxi Ganyue Expressway Co. Ltd), 000488 (Shandong Chenming Paper Group Co. Ltd), 000539 (Guangdong Electric Power Development Co. Ltd), 000541 (Foshan Electrical and Lighting Co. Ltd), 000550 (Jiangling Motors Co. Ltd), 000581 (Weifu High-Technology Co. Ltd), 000625 (Chongqing Changan Automobile Co. Ltd), 000709 (Tangshan Iron and Steel Co. Ltd), 000720 (Shandong Luneng Taishan Cable Co. Ltd), and 000778 (Xinxing Ductile Iron Pipes Co. Ltd).

The 23 stocks investigated in this work cover a variety of industry sectors including financials, real estate, conglomerates, metals \& nonmetals, electronics, utilities, IT, transportation, petrochemicals, paper \& printing and manufacturing. Our sample stocks were part of the 40 constituent stocks included in the Shenshen Stock Exchange Component Index in 2003~\cite{Zhou-2007-XXX}. More information of the data base and the market can be found in other relevant works \cite{Gu-Chen-Zhou-2007-EPJB,Jiang-Chen-Zhou-2008-PA,Mu-Chen-Kertesz-Zhou-2009-EPJB}.

\section{Defining large intraday price changes}
\label{S1:DefineEvents}

To study the limit order book dynamics around extreme events, we first have to give an exact definition of extreme events.  We generated a minute-to-minute dataset using mid-price of the last best bid and ask prices in every minute. In this paper, we continue to use a combined trigger to find the intraday events, as done in Ref.~\cite{Zawadowski-Kertesz-Andor-2004-PA,Zawadowski-Andor-Kertesz-2006-QF,Toth-Kertesz-Farmer-2009-EPJB}.

1.{\it Absolute filter}. This is used to find out the cumulative intraday price changes exceeding a given threshold value within a certain interval. In this paper, we choose 4\% and $\Delta t=60$ minutes as the threshold value and the length of time windows, respectively. Since the cumulative return $R_{\Delta t}(t)=[p(t)-p(t-\Delta t)]/p(t-\Delta t)\approx \ln[p(t)/p(t-\Delta t)]=r(t)$, for small $\Delta t$ we use the logarithmic return as an alternative quantity to the price change. However, some events located this way are just usual trades rather than extreme events because of the existence of intraday pattern in volatility. Moreover, a 4\% change may be an everyday event for a volatile stock, though this can be a signal of an extreme event for a less volatile stock.

2.{\it Relative filter}. We measure the average intraday volatility as another contrast. It means that the filter searches for intraday cumulative return exceeding 6 times the average volatility during the same time windows. Here, we utilize the sum of squared returns to define volatility of a certain time windows \cite{Bollen-Inder-2002-JEF,Andersen-Bollerslev-Diebold-Ebens-2001-JFE,Andersen-Bollerslev-Diebold-Labys-2001-JASA}
\begin{equation}
 v(t,\Delta t)=\sqrt{\sum_{\tau=t-\Delta t}^{t} r^2(\tau)}~,
 \label{Eq:v}
\end{equation}
where $\Delta t$ is length of time windows and $t$ is the end time of the time windows. This filter also has a weakness, because this method maybe only finds out the events happened in the period of low average volatility, but actual price changes are very small so that they are not true extreme events.

To be able to find out real intraday large events, we combine the two filters together and consider an event to be extreme if it passes both of them. Furthermore, we omit the first 5 minutes of trading day to get rid of opening effects. In order to investigate the intraday dynamics after events, we also have to omit the last 60 minutes of each trading day. Having found an event for the window size $\Delta t=60$ minutes, we shrink the window as long as the event still passes both filters. The time of the event, considered as $t=0$, is identified with the end of the window for the smallest $\Delta t$ satisfying these criteria.

\begin{figure}[htb]
\centering
\includegraphics[width=7cm]{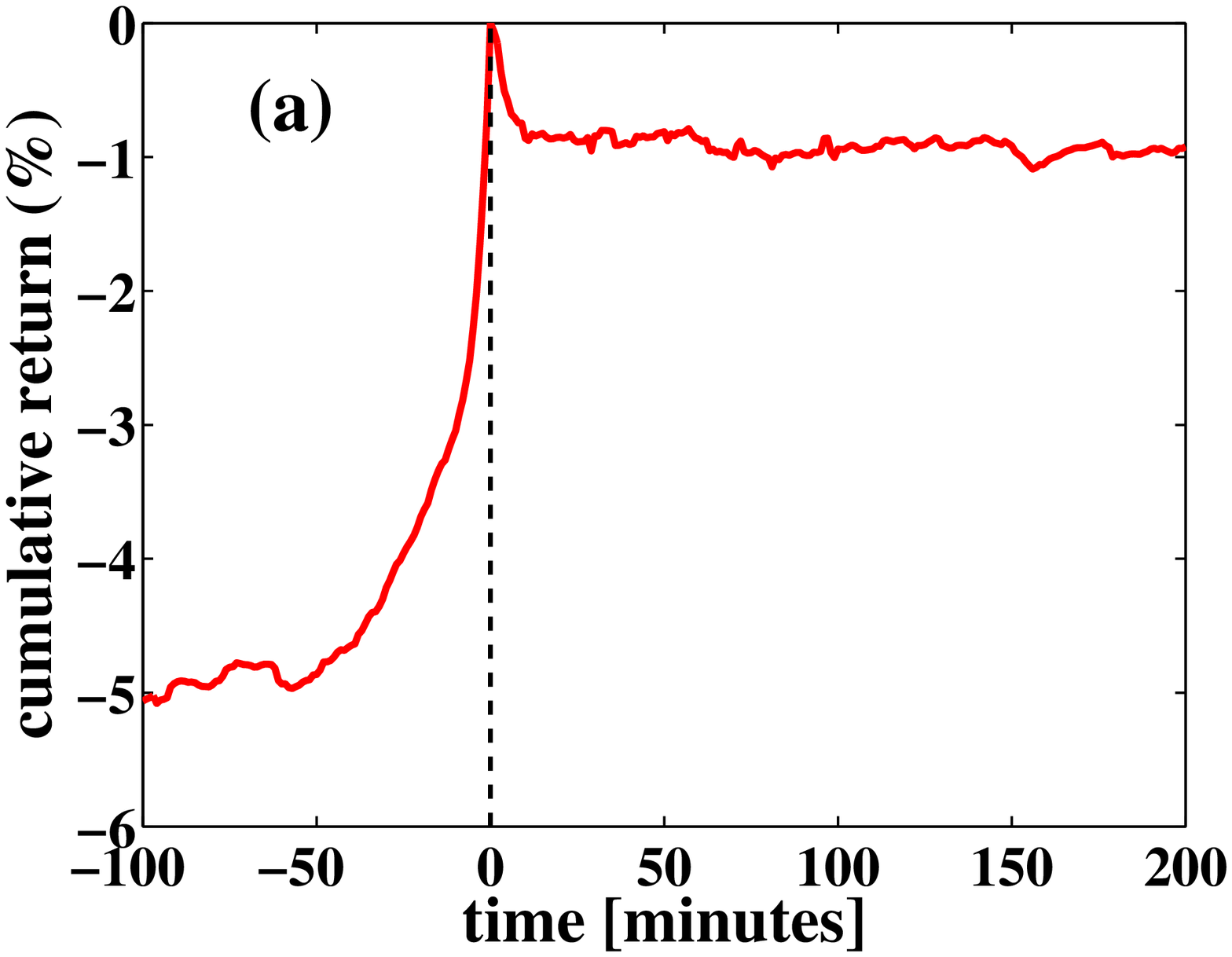}
\includegraphics[width=7cm]{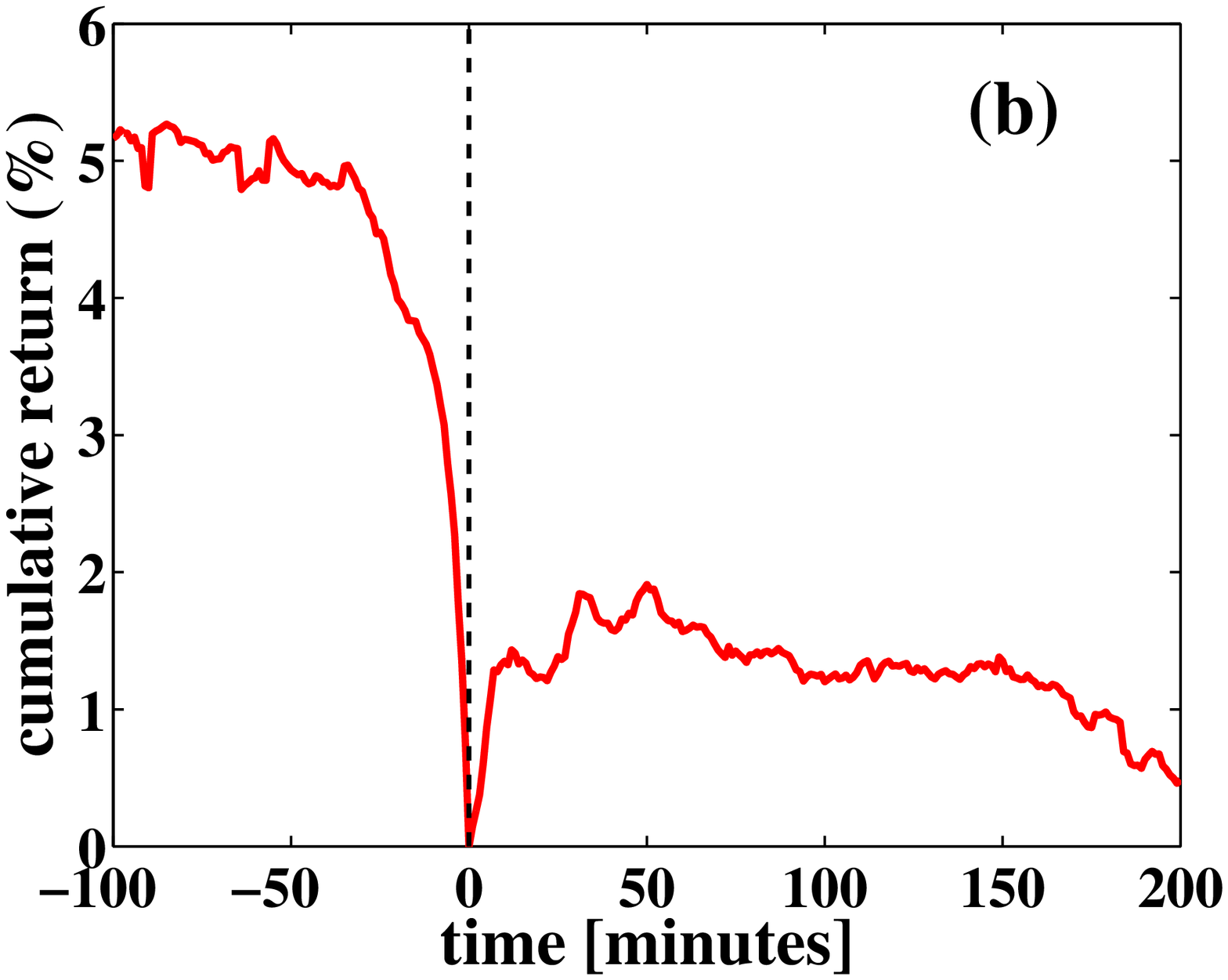}
\caption{\label{Fig:cumR} Cumulative return averaged over 131 price increases ((a) positive events) and 32 price decreases ((b) negative events), respectively. Minute 0 corresponds to the time when the price changes exceeds the both filters. Peak value happens at minute 0. Overreaction is obvious.}
\end{figure}

With the method above, we find 5450 events in 2003. In order to avoid computing a major event repeatedly, we only include the first events during a given day for a given stock in the average for convenience. Then 163 events are eligible, of which 131 events are positive events, others are negative events. Figure~\ref{Fig:cumR} gives average cumulative return for positive events and negative events, respectively. After averaging, the two curves are shifted respectively downward and upward so that the cumulative returns at $t=0$ are zero. An overreaction of around 1\% is found during the first 10 minutes for both, which is consistent with the results in Ref.~\cite{Zawadowski-Kertesz-Andor-2004-PA}, but in our case the peak is sharper. After overreaction, in the case of positive events the price relaxes to a rather stable value. This is not so  for negative events, though it should be mentioned that the much smaller number of negative events causes stronger fluctuations.

In the following, we will give some empirical results around the extreme events. Because eligible negative events are rare, we will place emphasis on the positive events in the following analysis. When we filter events, we only use the intraday data. However, when we investigate the dynamics over the events, the evolution is not restricted within one day.  Note that all events will have equal weight in the averaging procedure.

Before aggregating all the positive events, we have removed the usual dynamics from all investigated quantities. For each quantity $X(d,t')$, the intraday pattern $P(t')$ is determined as the reference, where $d$ identifies trading days and $t'=1,2,\cdots,240$ is the intraday time. The intraday pattern is removed from $X(d,t')$
\begin{equation}
 x(d,t') = X(d,t')/P(t')
 \label{Eq:x:d:t}
\end{equation}
for each trading day $d$. For each event $k$, the evolutionary trajectory $\{x_k(t):t=-100,-99,\cdots,-1,0,1,\cdots,200\}$ is extracted from $x(d,t')$, which contain 100 minutes before event $k$ and 200 minutes after event $k$. If the time distance of the event to the closing time is less than 200 minutes, we simply extend the time series to the next trading day, which is rational since the intraday pattern has been removed \cite{Zawadowski-Kertesz-Andor-2004-PA}. We then obtain the average of quantity $X$ for a group of events $\cal{K}$:
\begin{equation}
 x_{\cal{K}}(t) =
 \frac{1}{||{\cal{K}}||}\sum_{k\in{\cal{K}}}x_k(t),~~t=-100,\cdots,200,
 \label{Eq:X:K}
\end{equation}
where $||{\cal{K}}||$ is the number of events in group $\cal{K}$. Note that $t=0$ corresponds to the time when the extreme event happens.

\section{Dynamics of four basic quantities: absolute return, trading volume, spread and imbalance}
\label{S1:4measures}

In this section, we investigate the dynamics of four important quantities around positive and negative events. The first quantity is the 1-min absolute return. The second quantity is the total volume of market orders in one minute. Note that the volume has been adjusted for stock splitting. The third quantity is the bid-ask spread defined as the difference of best ask and best bid at the end of each minute. The fourth quantity is the buy imbalance for positive events defined as the ratio of total volume $V_b $ of buy market orders to the total volume $V_b + V_a$ of market orders in each minute \cite{Plerou-Gopikrishnan-Gabaix-Stanley-2002-PRE}:
\begin{equation}
 I = \frac{V_b}{V_b+V_s}.
 \label{Eq:I:Imbalance}
\end{equation}
The sell imbalance for negative events can be defined in an analogous way. All the events are partitioned into two groups, that is, 131 positive events and 32 negative events. We stress that we define the imbalance here for market orders \cite{Plerou-Gopikrishnan-Gabaix-Stanley-2002-PRE}, which is different from that in Ref.~\cite{Toth-Kertesz-Farmer-2009-EPJB} where both market orders and limit orders are included.

\subsection{Positive events}

\begin{figure}[htb]
\centering
\includegraphics[width=7cm]{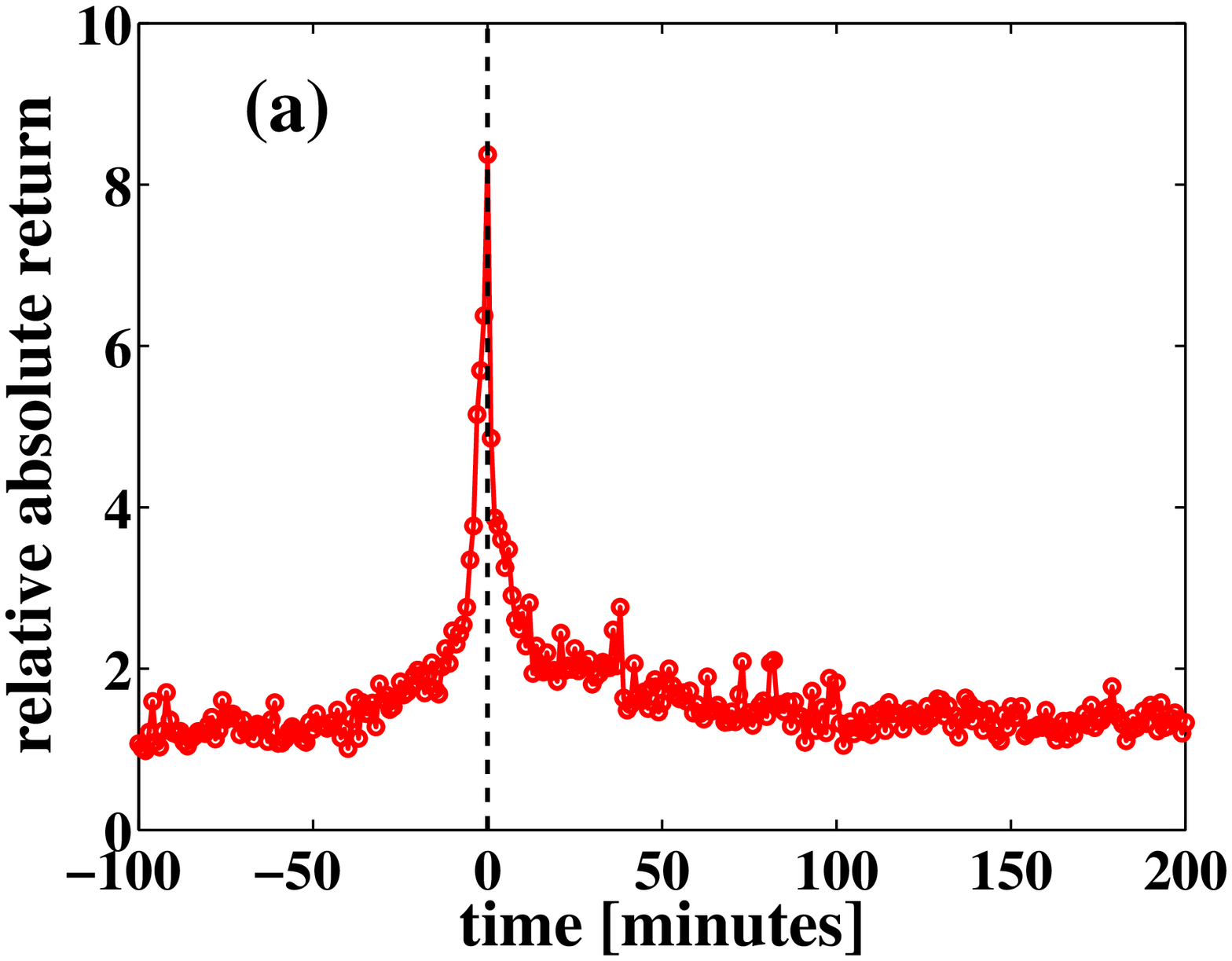}
\includegraphics[width=7cm]{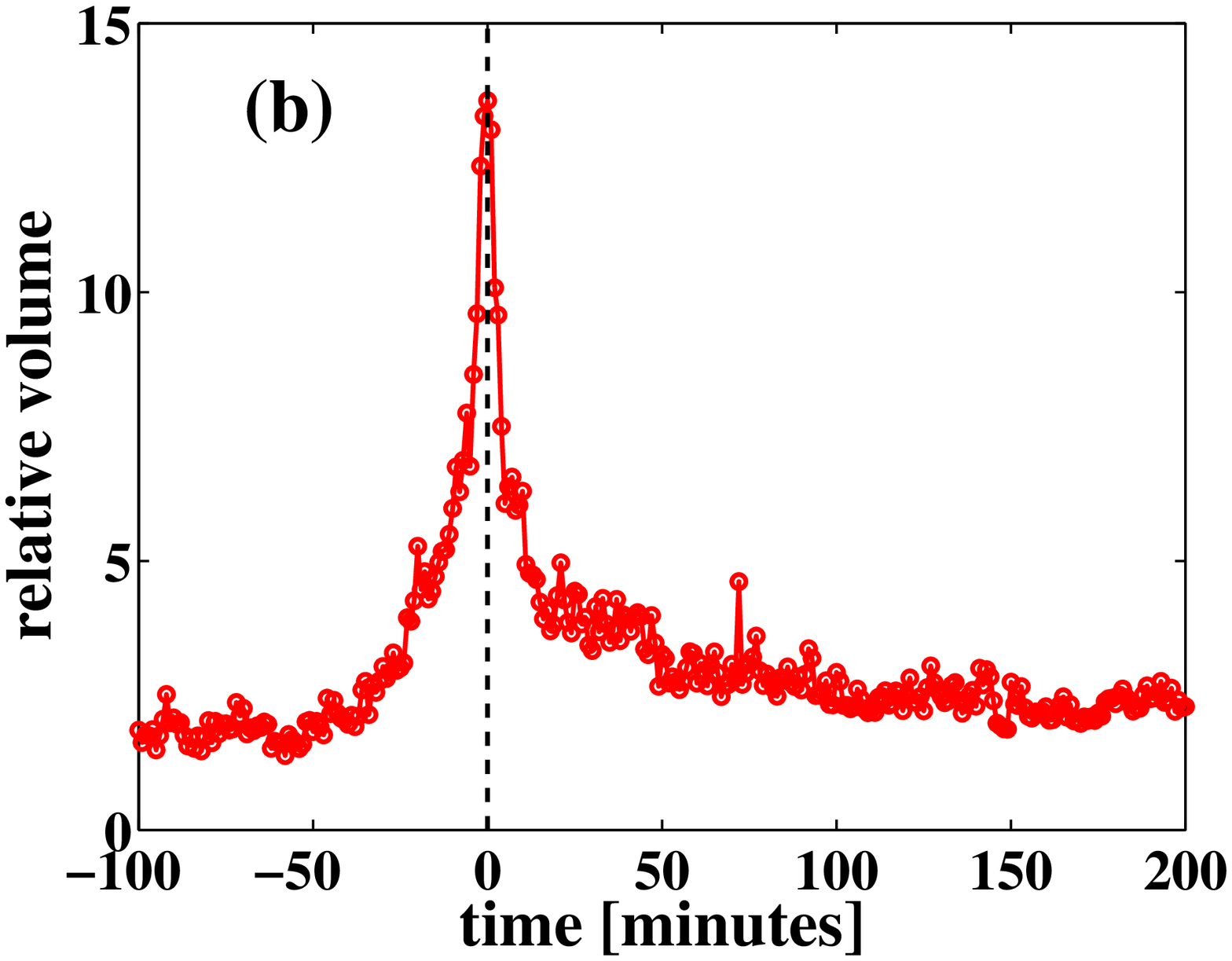}
\includegraphics[width=7cm]{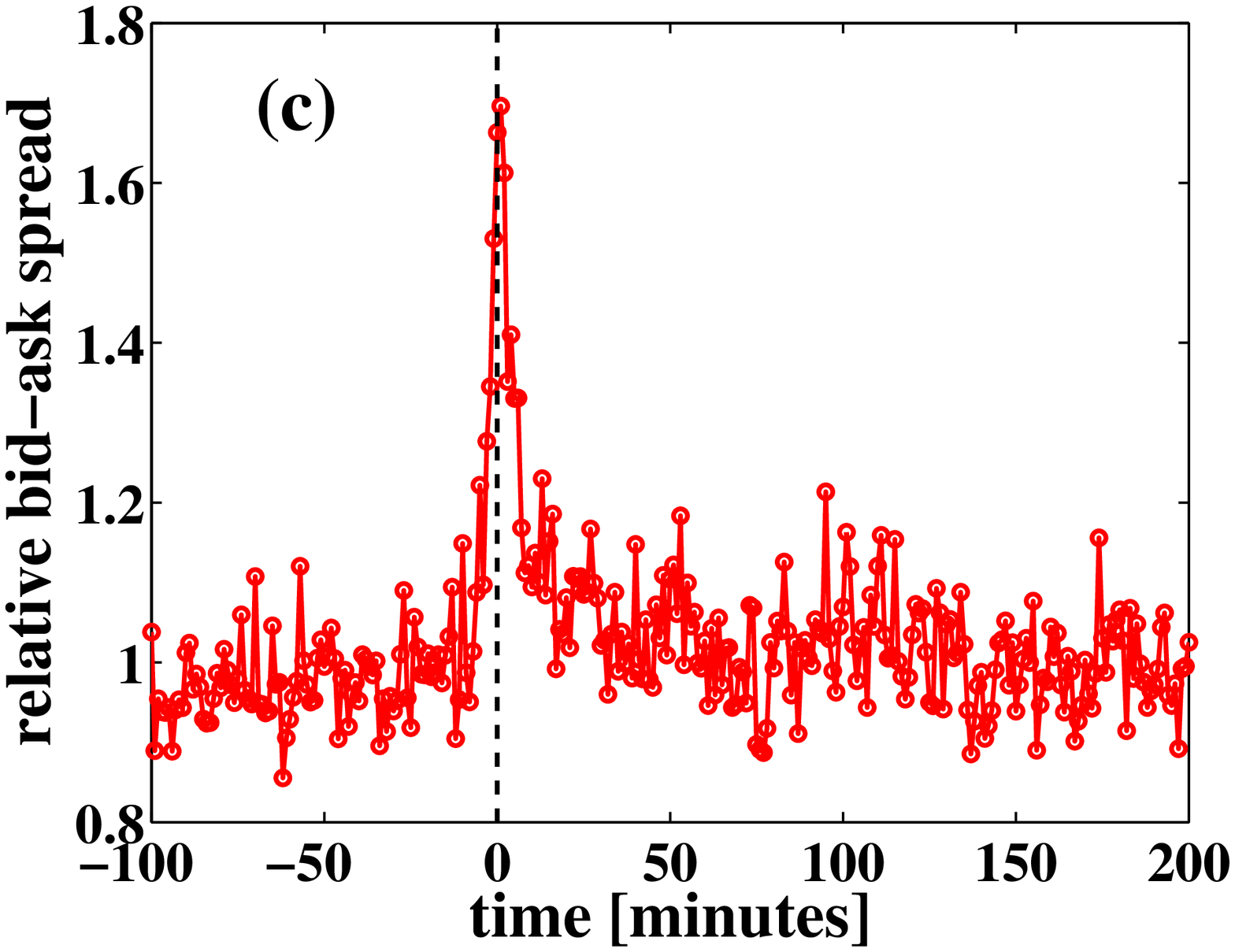}
\includegraphics[width=7cm]{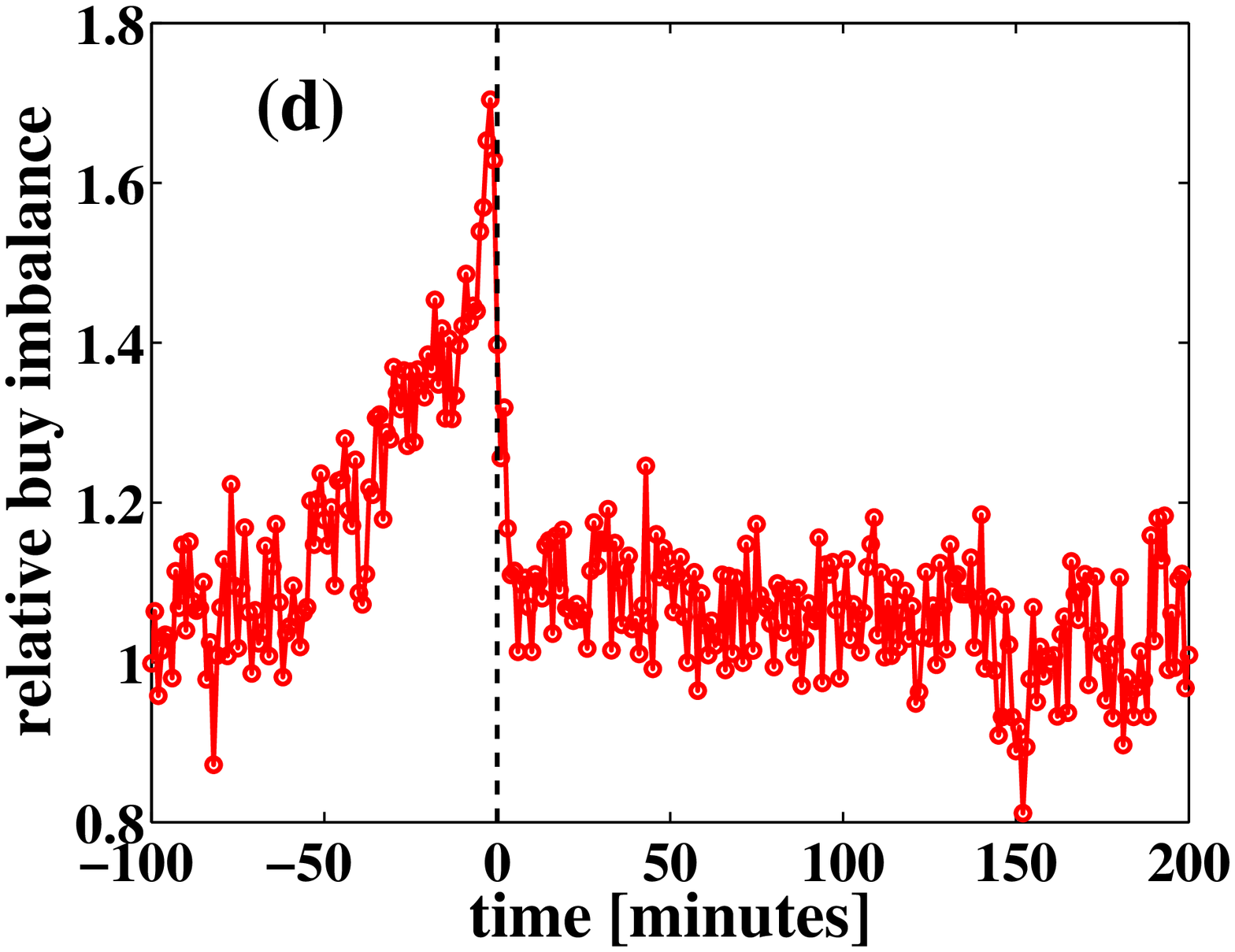}
\caption{\label{Fig:PositiveEvents} Dynamics of four variables around 131 positive events: (a) absolute return, (b) trading volume, (c) bid-ask spread, and (d) buy imbalance. The maximal values of bid-ask spread and buy imbalance locate at $t_{\max}= 1$ and $t_{\max}= -2$, respectively, while other two quantities have $t_{\max}= 0$. The vertical lines correspond to $t=0$.}
\end{figure}

Figure~\ref{Fig:PositiveEvents} gives the average evolution of the four quantities around 131 positive events. All the four plots share the same pattern with a sharp peak and an accumulation before the peak and a decay after that. The faster decay and very slow accumulation of the imbalance of volume is very different from the LSE case \cite{Toth-Kertesz-Farmer-2009-EPJB}. For absolute return, volume and imbalance, the pre-event accumulation is slower than the relaxation. In contrast, the post-event relaxation of the bid-ask spread is slower than its accumulation. The dynamics of the bid-ask spread in Fig.~\ref{Fig:PositiveEvents}c is very similar to the cases of NYSE and LSE but different from the NASDAQ case \cite{Zawadowski-Andor-Kertesz-2006-QF,Toth-Kertesz-Farmer-2009-EPJB}. An intriguing feature is that the imbalance of volume reaches its maximum at $t_{\max}=-2$, while the peaks of other three quantities happen at $t_{\max}=0$ for absolute return and volume and $t_{\max}=1$ for bid-ask spread respectively. This phenomenon in the imbalance dynamics is different from the LSE case in which $t_{\max}>0$ \cite{Toth-Kertesz-Farmer-2009-EPJB}, which might be due to the fact that we have a different definition of the imbalance.

Another intriguing feature of the buy imbalance dynamics around positive events is that the maximum of the relative buy imbalance is about 1.70. We found empirically that the unconditional average of the buy imbalance is 0.47, which is reminiscent of the fact that the Chinese stock market was bear from mid-2001 to 2005 \cite{Zhou-Sornette-2004a-PA}. On average, about 79.9\% of the total volume of executed orders was placed on the buy side in the preceding minute before extreme positive events. It is natural to assume that there is a monotonic relationship between the imbalance and the price difference. Accordingly, we observe the same kind of asymmetry in the shape of the pre- and post-event graphs of these quantities. Note that the difference to Fig. 3a of \cite{Toth-Kertesz-Farmer-2009-EPJB} comes from the different definitions of the imbalance.

\subsection{Negative events}

\begin{figure}[htb]
\centering
\includegraphics[width=7cm]{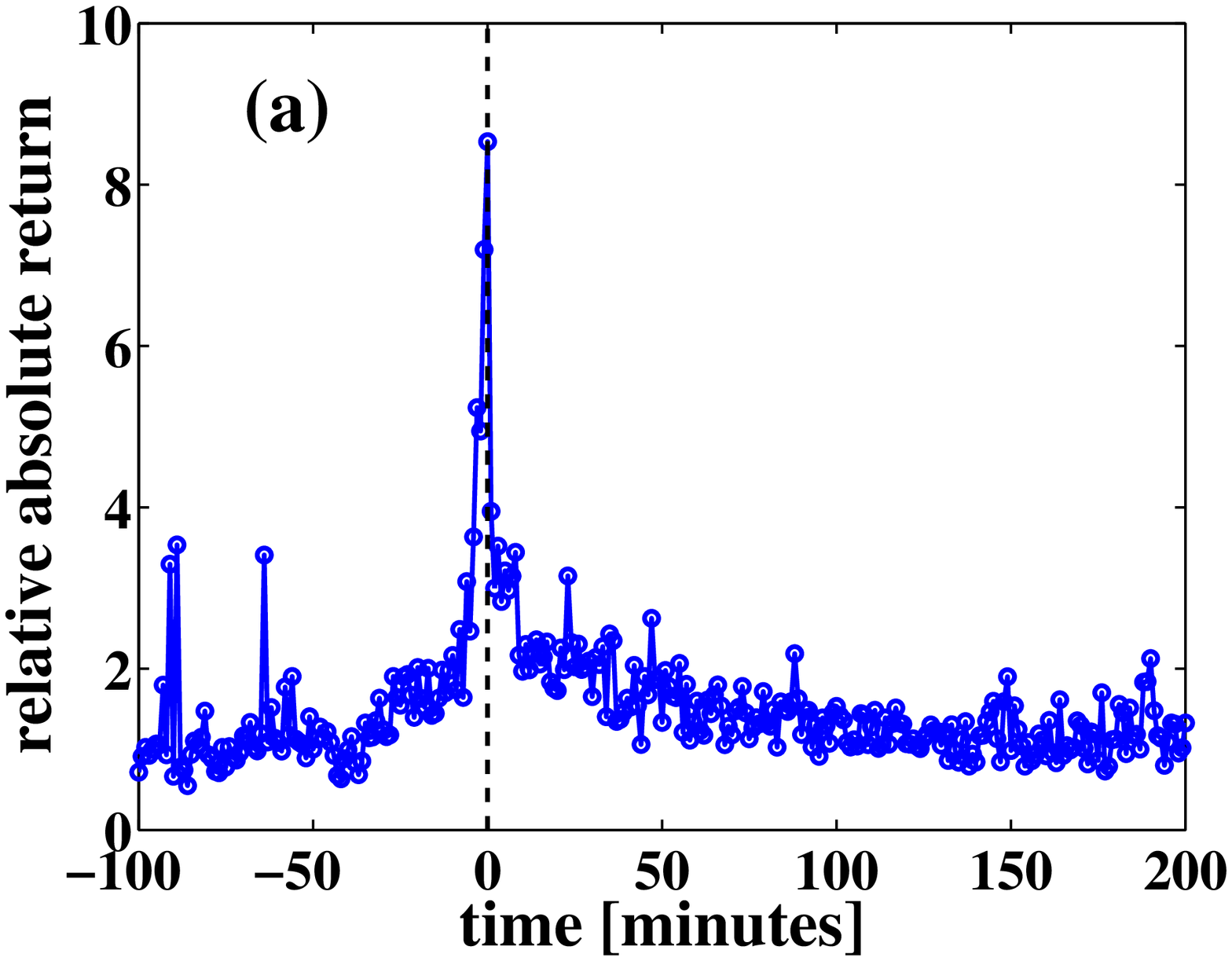}
\includegraphics[width=7cm]{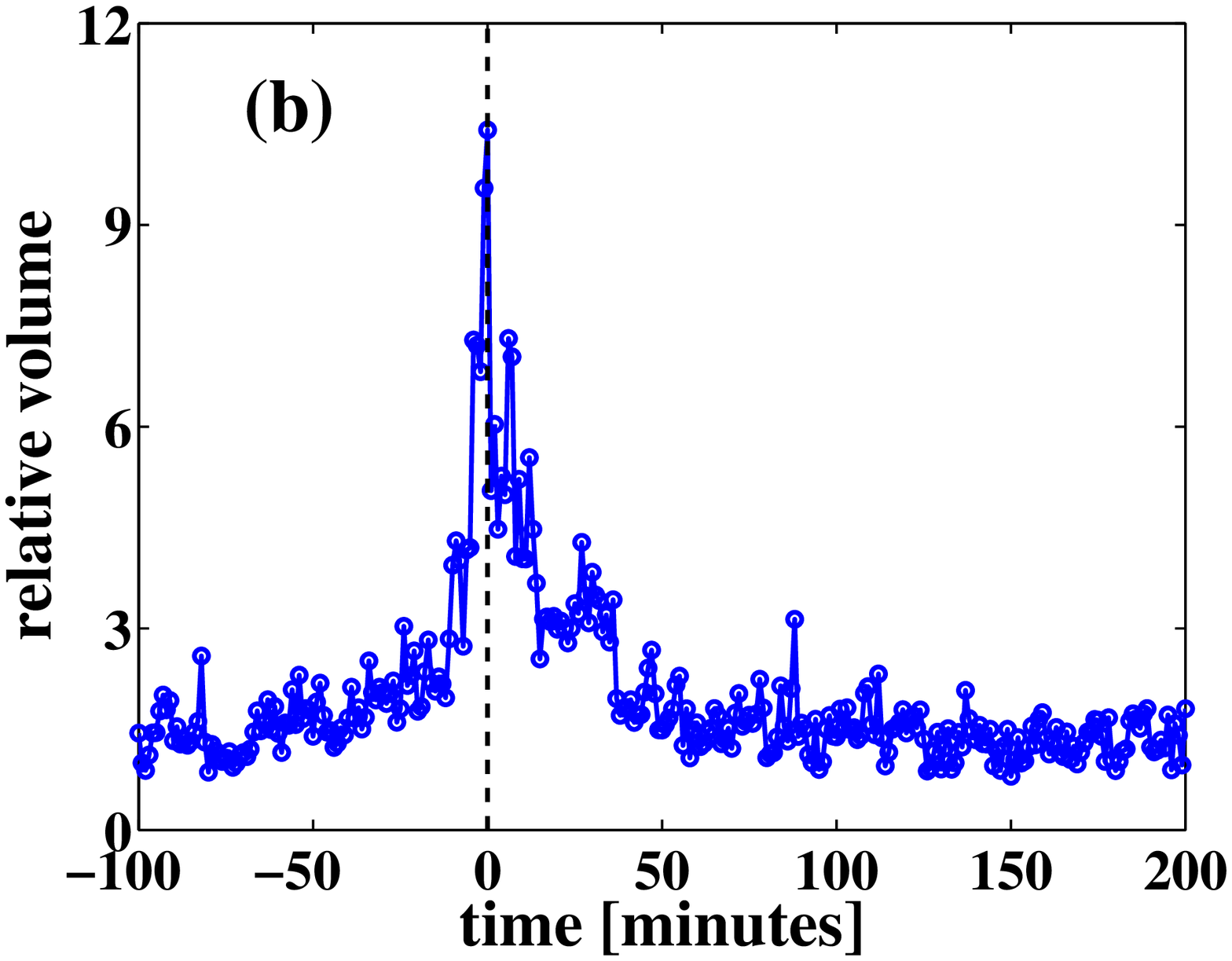}
\includegraphics[width=7cm]{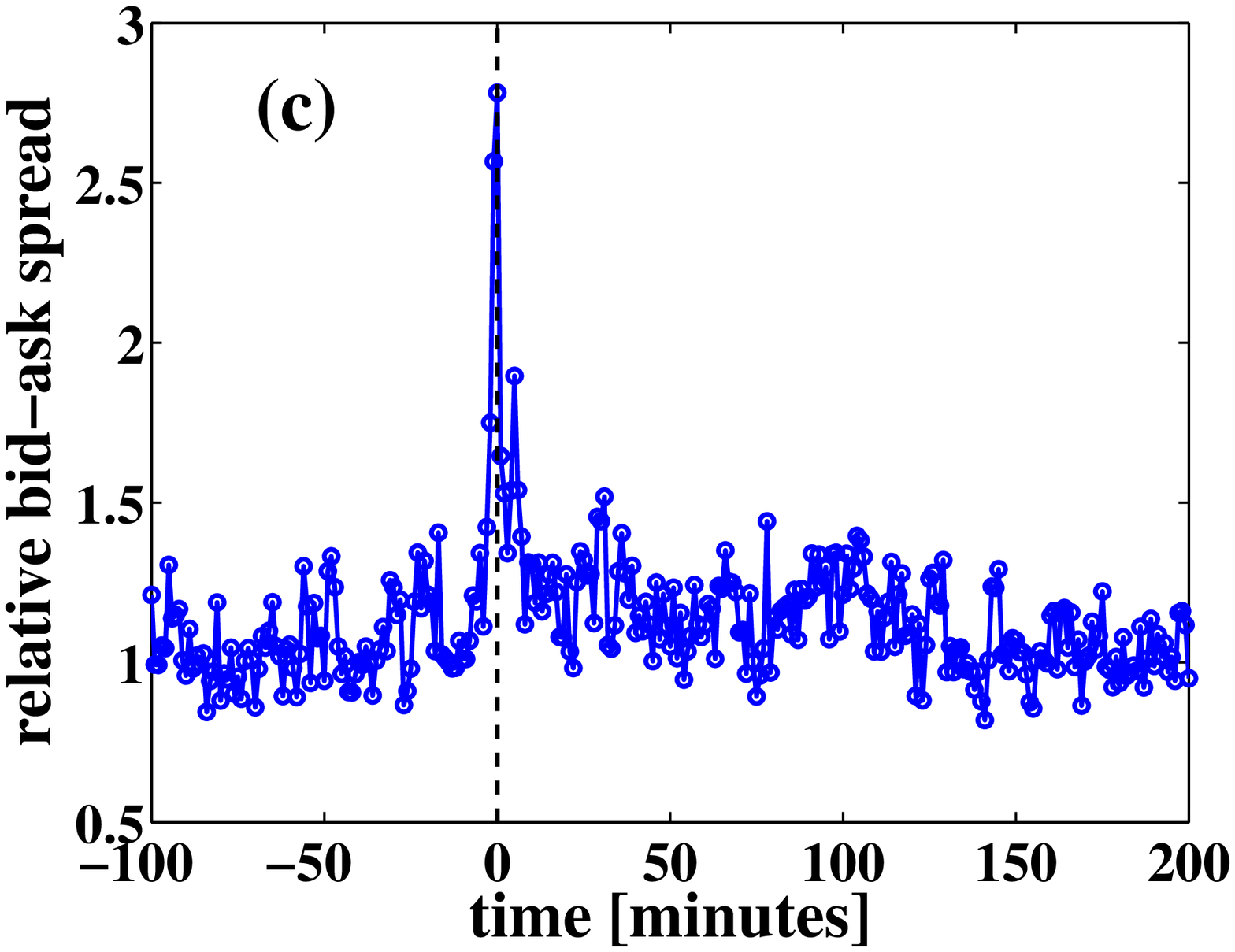}
\includegraphics[width=7cm]{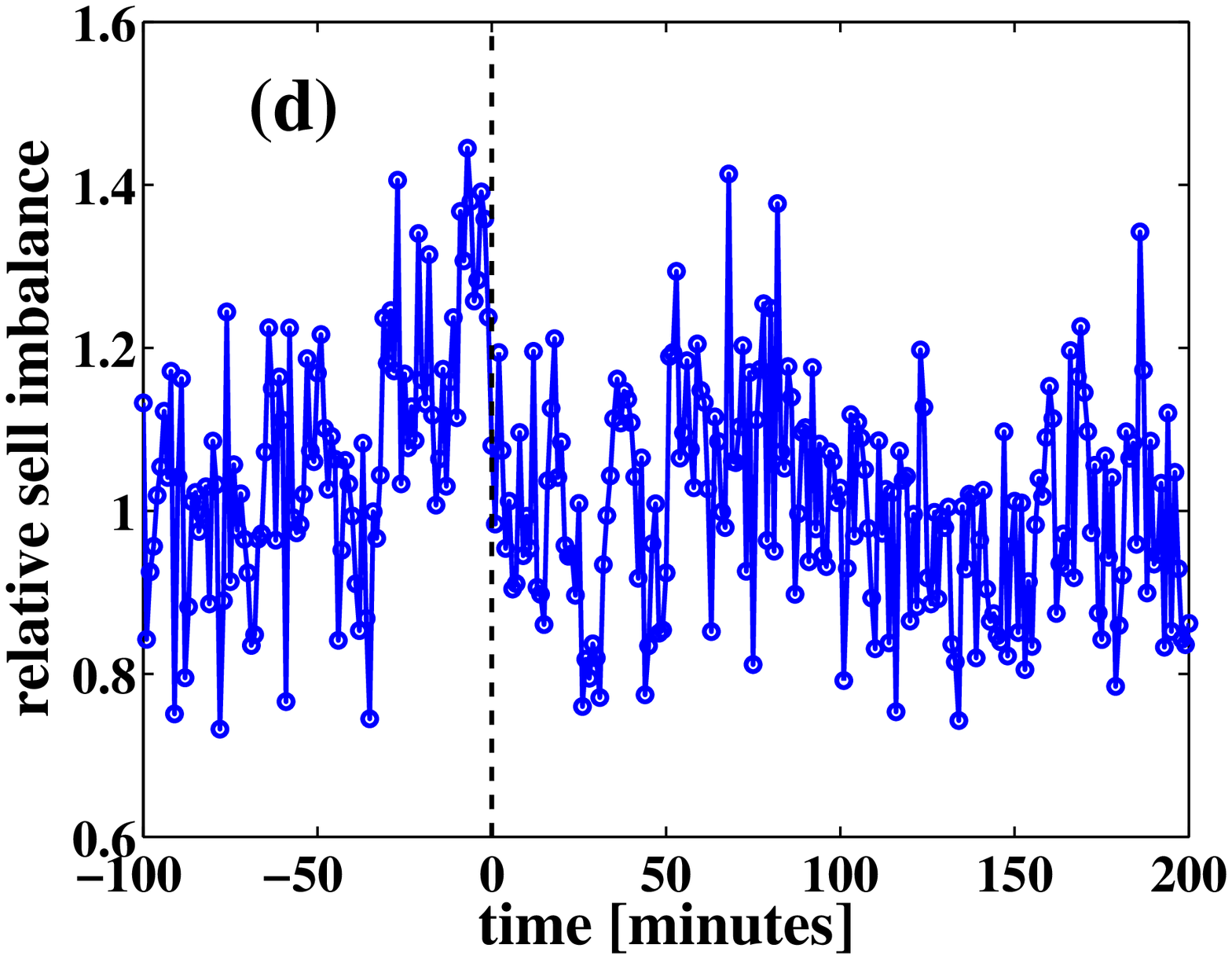}
\caption{\label{Fig:NegativeEvents} Dynamics of four variables around 32 negative events: (a) absolute return, (b) trading volume, (c) bid-ask spread, and (d) sell imbalance. The maxima of the quantities are $t_{\max}= 0$, except for the imbalance, where the data are too noisy to enable the identification of a proper maximum. The vertical lines correspond to $t=0$.}
\end{figure}

Figure~\ref{Fig:NegativeEvents} gives the average evolution of the four quantities around 32 negative events. We find that the four temporal evolutions around negative events are qualitatively the same as those around positive events. It is not surprising that the curves for negative events have larger fluctuations than the curves for positive events, since there are much more positive events recognized from the data. All the four trajectories share the same pattern with a sharp peak and an accumulation before the peak and a decay after that. For absolute return, volume and imbalance, the accumulation is slower than the relaxation. Again, the dynamics of the bid-ask spread in Fig.~\ref{Fig:PositiveEvents}c is similar to the cases of NYSE and LSE but different from the NASDAQ case \cite{Zawadowski-Andor-Kertesz-2006-QF,Toth-Kertesz-Farmer-2009-EPJB}. We observe that the peaks of absolute return, volume and bid-ask spread happen at $t_{\max}=0$. Although the imbalance curve is very noisy, we are still able to identify the peak at $t_{\max}=-7$, which is also different from the LSE case where $t_{\max}>0$ \cite{Toth-Kertesz-Farmer-2009-EPJB}.

From Fig.~\ref{Fig:NegativeEvents}d, the maximum sell imbalance around negative evens is about 1.4. In addition, the unconditional average sell imbalance is $I_0=0.53$. It follows that, on average, about 74\% of the total volume of executed orders was placed on the sell side in the preceding minute before extreme negative events.

\subsection{Power-law relaxation}
\label{S1:4measures_PL}

\begin{figure}[htb]
\centering
\includegraphics[width=7cm]{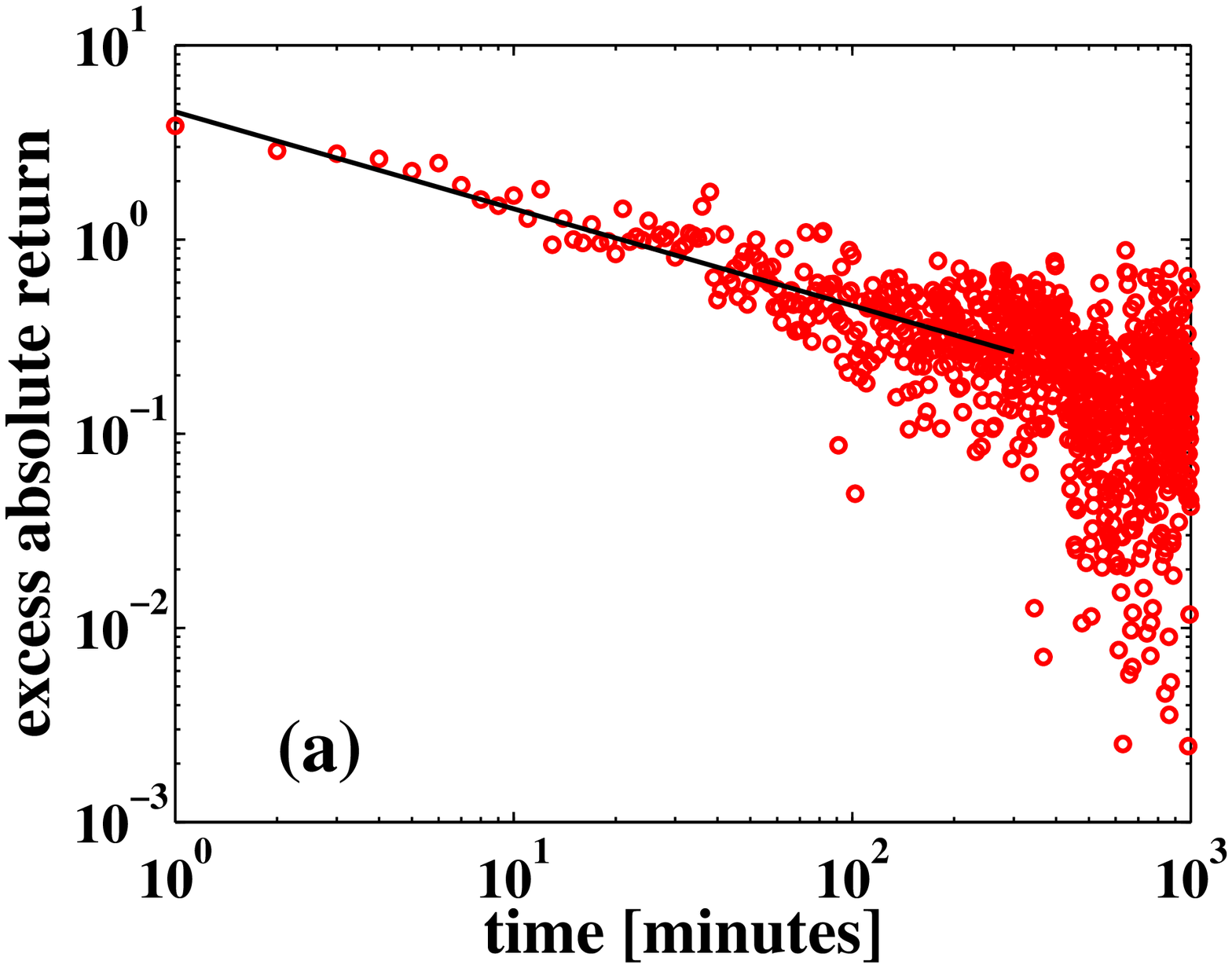}
\includegraphics[width=7cm]{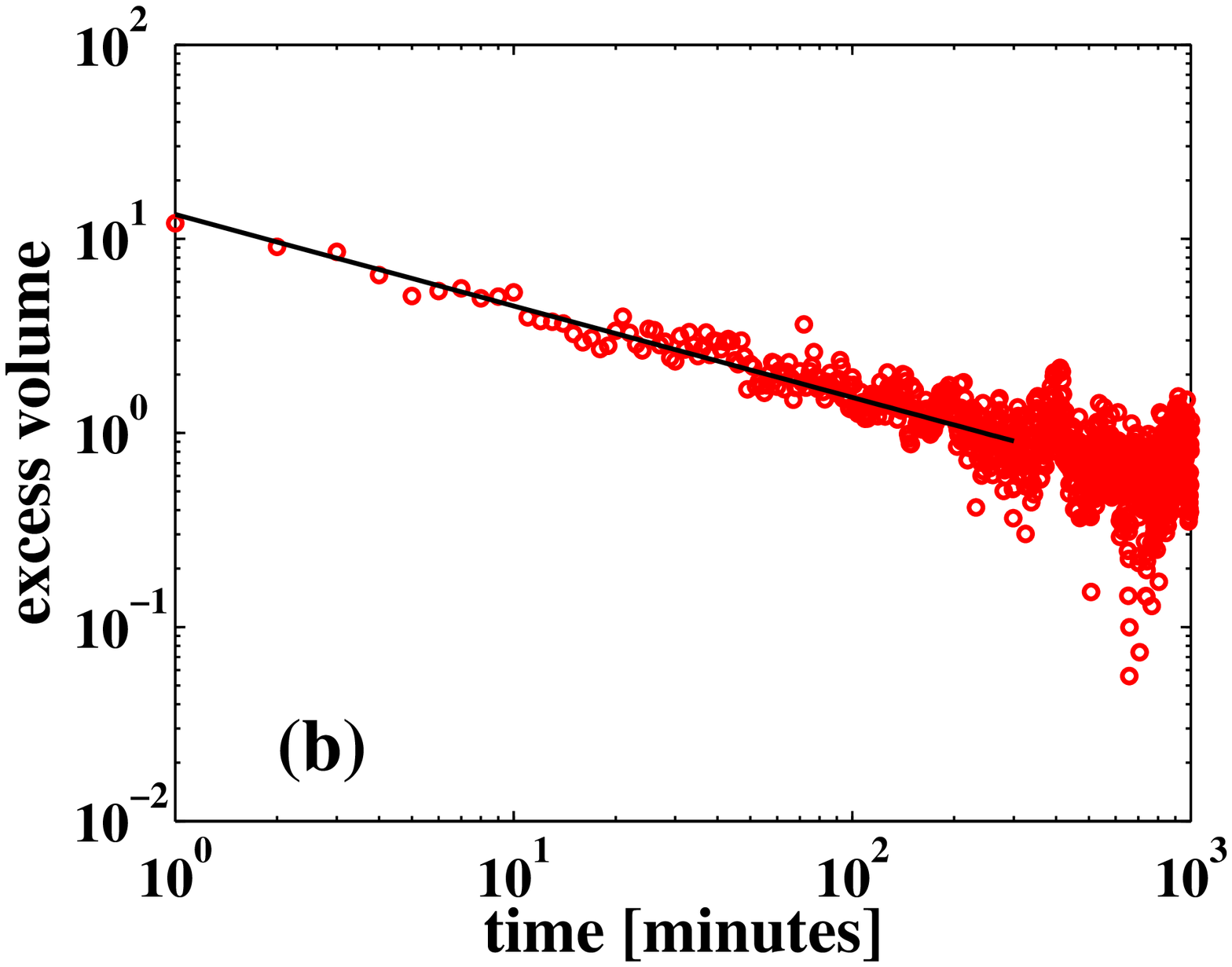}
\includegraphics[width=7cm]{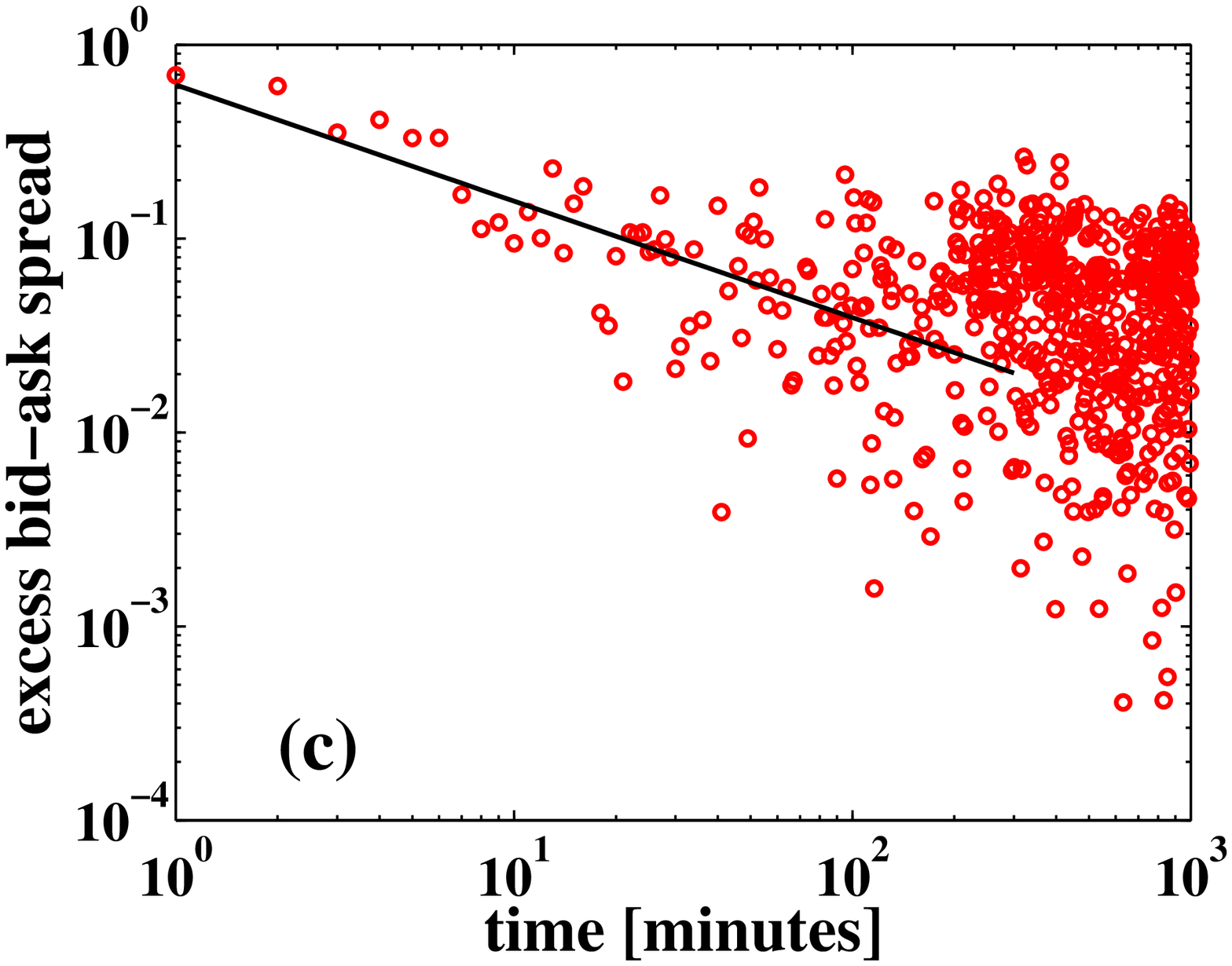}
\includegraphics[width=7cm]{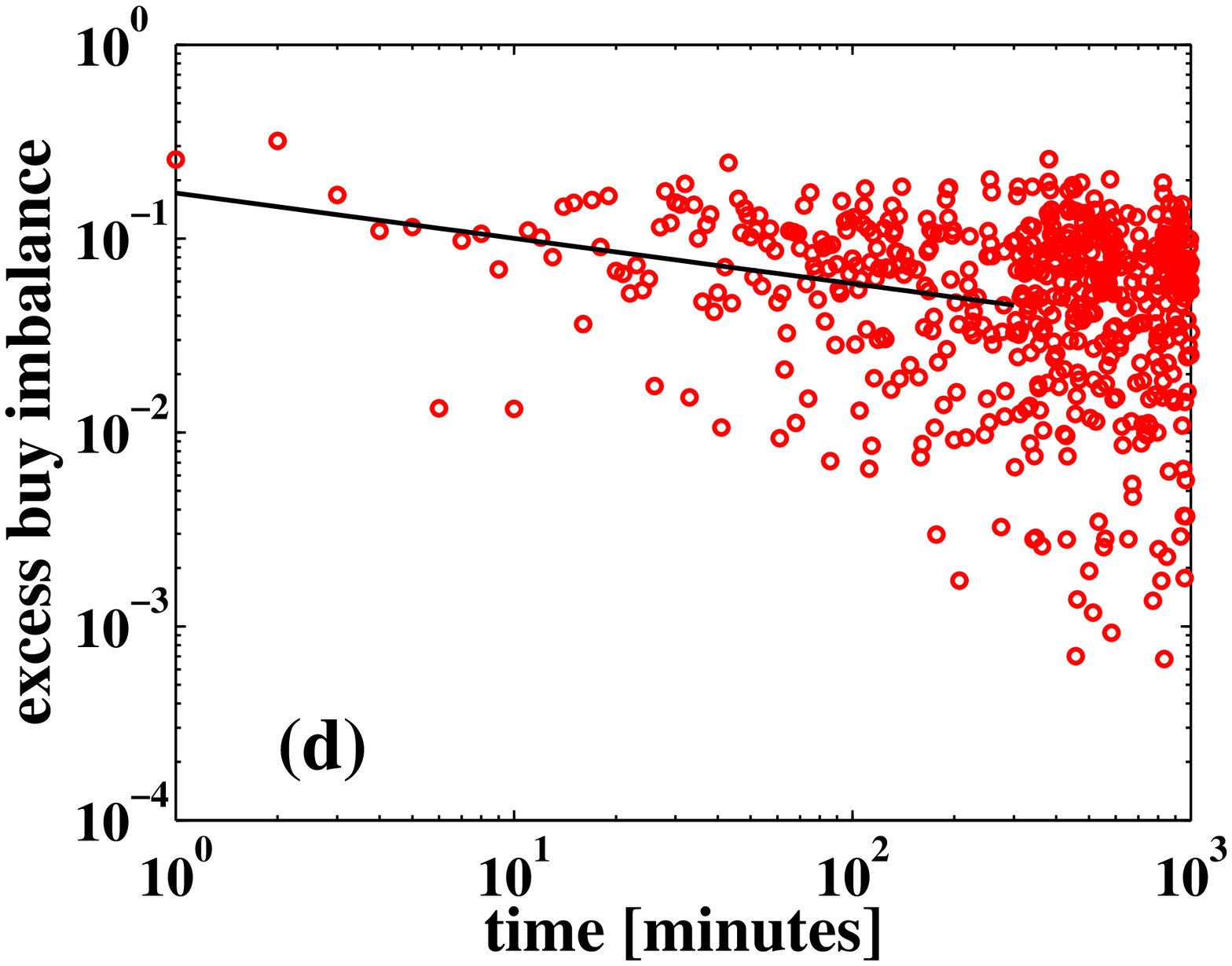}
\caption{\label{Fig:PositiveEvents:PL} Power-law relaxation of excess variables for the absolute mid-price return, the trading volume, the bid-ask spread and the buy imbalance after positive extreme events on log-log scales with power-law fits. The relaxation exponents $\alpha$ are $0.50\pm0.03$, $0.47\pm0.02$, $0.60\pm0.09$ and $0.23\pm0.06$, respectively.}
\end{figure}

To study the relaxations to the normal value, we plot the excess variables defined as the difference between the actual value and the value in normal periods,
\begin{equation}
 x_{\rm{ex}}(t)=x(t)-x_0=x(t)-1,
 \label{Eq:xt:EX:PL}
\end{equation}
where the value in normal periods $x_0$ is 1 by definition. Figure \ref{Fig:PositiveEvents:PL} illustrates the relaxation of the excess variables around positive events on log-log scales. All the curves exhibit power-law behaviors
\begin{equation}
 x_{\rm{ex}}(t) \sim t^{-\alpha},
 \label{Eq:xt:PL}
\end{equation}
which has the same form as those for western mature stock markets \cite{Zawadowski-Kertesz-Andor-2004-PA,Zawadowski-Andor-Kertesz-2006-QF,Toth-Kertesz-Farmer-2009-EPJB,Ponzi-Lillo-Mantegna-2009-PRE}. The power-law relaxation is more significant with less fluctuations and longer scaling ranges for absolute return and volume than for bid-ask spread and volume imbalance.

The estimates of the $\alpha$ values can be obtained through the linear least-squares regression of $\ln x$ against $\ln t$, and we have $\alpha_{\rm{AR}}=0.50\pm0.03$ for absolute return, $\alpha_{\rm{V}}=0.47\pm0.02$ for volume, $\alpha_{\rm{BAS}}=0.60\pm0.09$ for bid-ask spread, and $\alpha_{\rm{Imb}}=0.23\pm0.06$ for volume imbalance. Similarly, there are also power-law decay for negative events and we find that $\alpha_{\rm{AR}}=0.53\pm0.02$ for absolute return, $\alpha_{\rm{V}}=0.64\pm0.08$ for volume, $\alpha_{\rm{BAS}}=0.54\pm0.11$ for bid-ask spread, while for the volume imbalance the fit is too vague. In both cases of positive and negative events, the fitting was performed on the range $[1,300]$ for absolute return, volume, bid-ask spread and imbalance. All the resulting exponents are listed in Table \ref{Tb:4measures:PL}.

\begin{table}[htb]
 \caption{\label{Tb:4measures:PL} Relaxation exponents of different variables.}
 \medskip
 \centering
 \begin{tabular}{lcccccccccccc}
  \hline\hline
   \multirow{3}*[1.5mm]{Variable} & \multicolumn{2}{@{\extracolsep\fill}c}{Exponents} \\
\cline{2-3}
  & Positive events & Negative events \\
  \hline
  Absolute return  &  $0.50\pm0.03$ & $0.53\pm0.02$ \\%
  Volume           &  $0.47\pm0.02$ & $0.64\pm0.08$ \\%
  Bid-ask spread   &  $0.60\pm0.09$ & $0.54\pm0.11$ \\%
  Volume imbalance &  $0.23\pm0.06$ & --            \\%
  \hline\hline
 \end{tabular}
\end{table}

The relaxation of absolute return for the SZSE stocks is faster than that for the western stocks in which the exponent $\alpha_{\rm{AR}}=0.39\pm0.01$ (negative events) and $\alpha_{\rm{AR}}=0.34\pm0.01$ (positive events) for NYSE stocks, and $\alpha_{\rm{AR}}=0.43\pm0.01$ (negative events) and $\alpha=0.44\pm0.01$ (positive) for NASDAQ stocks \cite{Zawadowski-Kertesz-Andor-2004-PA}, $\alpha=0.25\sim0.40$ for NYSE stocks and $\alpha=0.32\sim0.41$ for NASDAQ stocks \cite{Zawadowski-Andor-Kertesz-2006-QF}, and $\alpha=0.38\pm0.01$ (all events) for LSE stocks \cite{Toth-Kertesz-Farmer-2009-EPJB}. In addition, the relaxation exponents of the bid-ask spread of SZSE stocks are also different from that of LSE stocks where $\alpha=0.38\pm0.03$ \cite{Toth-Kertesz-Farmer-2009-EPJB,Ponzi-Lillo-Mantegna-2009-PRE}. A comparison of the resulting exponents is illustrated in Table \ref{Tb:Comparion:PL}.

\begin{table}[htb]
 \caption{\label{Tb:Comparion:PL} Relaxation exponents of absolute return for different stock markets.}
 \medskip
 \centering
 \begin{tabular}{lccccccccccccc}
  \hline\hline
   \multirow{3}*[1.5mm]{Stock markets} && \multicolumn{2}{@{\extracolsep\fill}c}{Exponents} \\
\cline{3-4}
            && Positive events & Negative events   \\%
  \hline
  SZSE      &&  $0.50\pm0.03$  & $0.53\pm0.02$     \\%
  NYSE\cite{Zawadowski-Kertesz-Andor-2004-PA}
            &&  $0.34\pm0.01$  & $0.39\pm0.01$     \\%
  NASDAQ\cite{Zawadowski-Kertesz-Andor-2004-PA}
            &&  $0.44\pm0.01$  & $0.43\pm0.01$     \\%
  NYSE\cite{Zawadowski-Andor-Kertesz-2006-QF}
            &&  $0.25\sim0.40$ & $0.25\sim0.38$    \\%
  NASDAQ\cite{Zawadowski-Andor-Kertesz-2006-QF}
            &&  $0.35\sim0.41$ & $0.32\sim0.40$    \\%
  LSE\cite{Toth-Kertesz-Farmer-2009-EPJB}
            &&  --             & $0.37\pm0.01$     \\%
  \hline\hline
 \end{tabular}
\end{table}

\section{Volume dynamics of buy and sell market orders}
\label{S1:Buy:Sell}

In this section we investigate the flow dynamics of buy and sell market orders around extreme events. The intraday pattern is determined for each quantity, which has been removed according to Eq.~(\ref{Eq:x:d:t}). The events are divided into two groups, one containing all the 131 positive events and the other including all the 32 negative events. In this section we consider only the positive events. The behavior around negative ones is similar (with exchanged roles of buyer and seller initiated orders), though the fluctuations are considerably larger due to the much smaller number of negtive events.

\subsection{Pre-event and post-event dynamics}

\begin{figure}[htb]
\centering
\includegraphics[width=7cm]{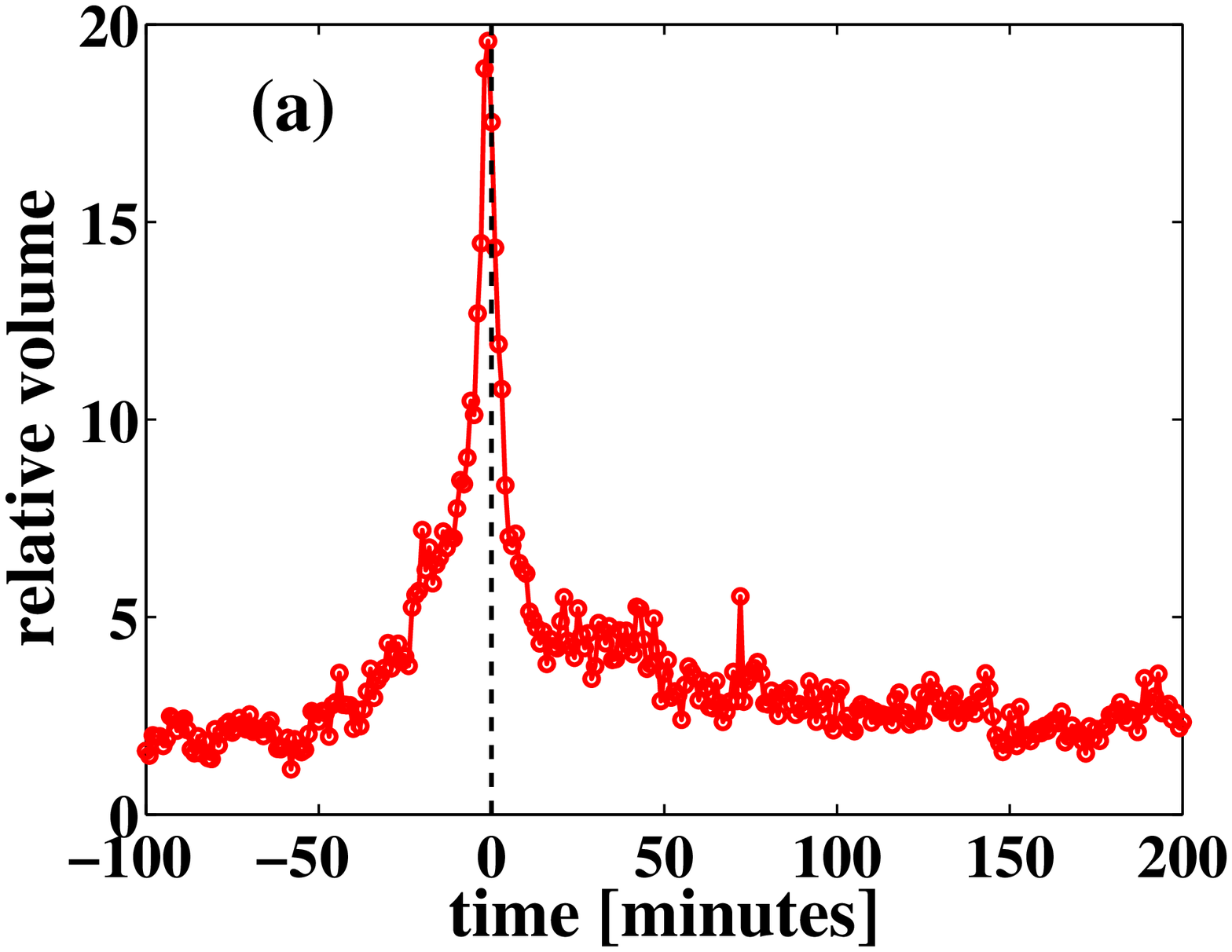}
\includegraphics[width=7cm]{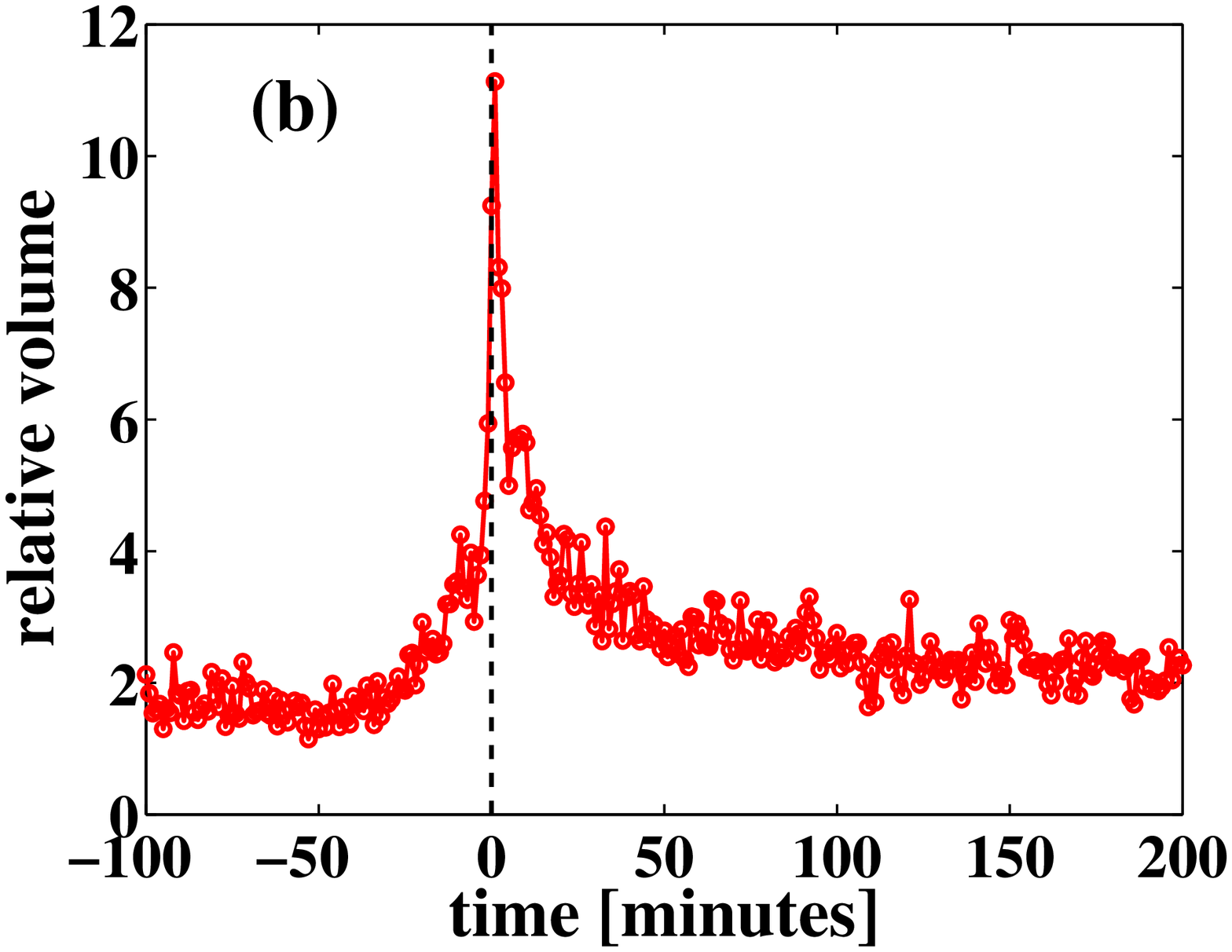}
\caption{\label{Fig:Direction} Dynamics of relative volume of buy market orders and sell market orders around 131 positive events. The moment that the volume reaches its maximum is (a) $t_{\max}=-1$ for buys around positive events and (b) $t_{\max}=1$ for sells around positive events.}
\end{figure}

Figure \ref{Fig:Direction} shows the dynamics of the two quantities. Roughly speaking, the shapes of the dynamics are similar to that for volume shown in Fig.~\ref{Fig:PositiveEvents}b and Fig.~\ref{Fig:NegativeEvents}b. Also, the two trajectories around negative events are noisier than positive events. The buy volume around positive events reaches its maximum at $t_{\max}=-1$, which is about 20 times its average value in regular moments. The sell volume around positive events reaches its maximum at $t_{\max}=1$, which is about 12 times its average value in regular moments. The buy volume around negative events reaches its maximum at $t_{\max}=0$, which is about 8 times its average value in regular moments. The sell volume around positive events reaches its maximum at $t_{\max}=-1$, which is about 13 times its average value in regular moments. Around positive events, the volume of buy market orders are larger than sell market orders, which implies a persistent positive buy imbalance of volumes driving the price up. These observation is consistent with the positivity of buy imbalance in Fig.~\ref{Fig:PositiveEvents}d and sell imbalance in Fig.~\ref{Fig:NegativeEvents}d and in agreement with Fig.~\ref{Fig:cumR}.

Another very interesting feature is the significant difference in the structure of the peaks for buy orders and sell orders. This can be intuitively explained by a propagation chain of order flows. Due to an endogenous or exogenous process the price starts to rise. The hope that the process will lead to a new, higher average value of the stock drives the buyers to place an increasing number of orders. With a little delay the number of sell market orders also increases indicating the intention to realize the profit due to the price change. The correlation between order size and liquidity was verified empirically \cite{Farmer-Gillemot-Lillo-Mike-Sen-2004-QF}. Before the peak at $t=-1$, the aggregate price impact of buy market orders are greater than sell market orders and the price increases continually. This trend reverses at time $t=0$ when the price impact of buy executed orders is identical to sell executed orders. After $t=0$, the price impact of sell executed orders dominates and the price drops. The situation around negative events is very similar.

\subsection{Power-law relaxation}

\begin{figure}[htb]
\centering
\includegraphics[width=8cm]{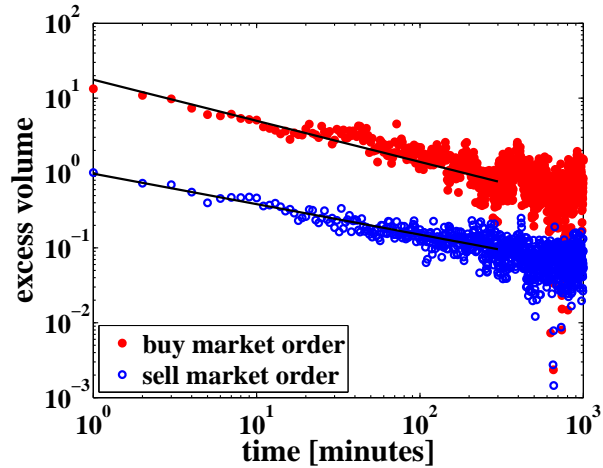}
\caption{\label{Fig:Direction:PL}  Power-law relaxation of the excess volume of buy market orders and sell market orders after positive events on log-log scales. The curves have been shifted vertically for clarity.}
\end{figure}

Figure \ref{Fig:Direction:PL} illustrates the post-event relaxation of the excess volume of buy market orders and sell market orders around positive events on log-log scales, respectively. All the curves exhibit power-law behaviors. The scaling ranges span more than two orders of magnitude. The relaxation exponent is $\alpha_{\rm{buy}} = 0.55\pm0.03$ for buy market orders and $\alpha_{\rm{sell}} = 0.41\pm0.02$ for sell market orders around positive events. The fitting was performed on the range $[1,300]$.

\section{Volume dynamics of four types of orders with different aggressiveness}
\label{S1:Agg}

We have shown that order direction (buy or sell) can be used to classify the dynamics of relative volume of incoming market orders. In this section, we investigate further the dynamics of different types of orders around extreme events by taking into to
account order aggressiveness. Orders can be classified into different types according to their aggressiveness \cite{Biais-Hillion-Spatt-1995-JF}. Here, we consider four types of orders on each side of the order flow, that is, partially filled market orders, filled market orders, limit orders and canceled orders. More rigorously, ``market orders'' and ``limit orders'' should be termed ``effective market orders'' and ``effective limit orders'' \cite{Smith-Farmer-Gillemot-Krishnamurthy-2003-QF}, since there were no market orders permitted to submit in the SZSE in 2003 \cite{Gu-Chen-Zhou-2007-EPJB}. The differentiation of partially filled orders and filled orders is validated due to the fact that they have very different behavior of price impact \cite{Zhou-2007-XXX}

We study the dynamics of volumes of four types of buy orders and sell orders around the extreme events. Therefore, there are eight quantities. The intraday pattern is determined for each quantity, which has been removed according to Eq.~(\ref{Eq:x:d:t}). Again, the events are divided into two groups, one containing all the 131 positive events and the other including 32 negative events. In addition, we also study the dynamics of numbers of buy orders and sell orders around the extreme event. We find that the results are quite similar with minor differences.

\begin{figure}[htb]
\centering
\includegraphics[width=7cm]{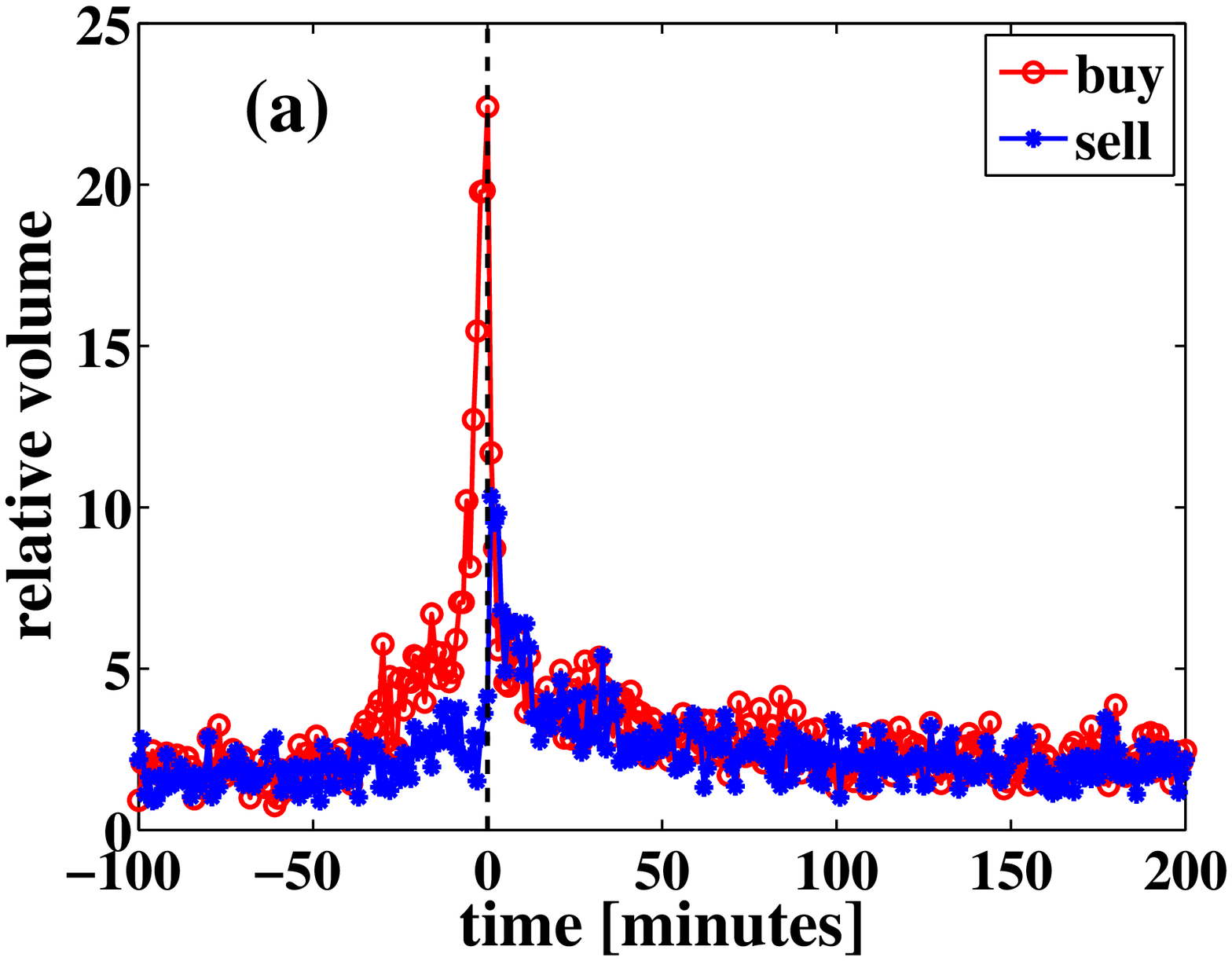}
\includegraphics[width=7cm]{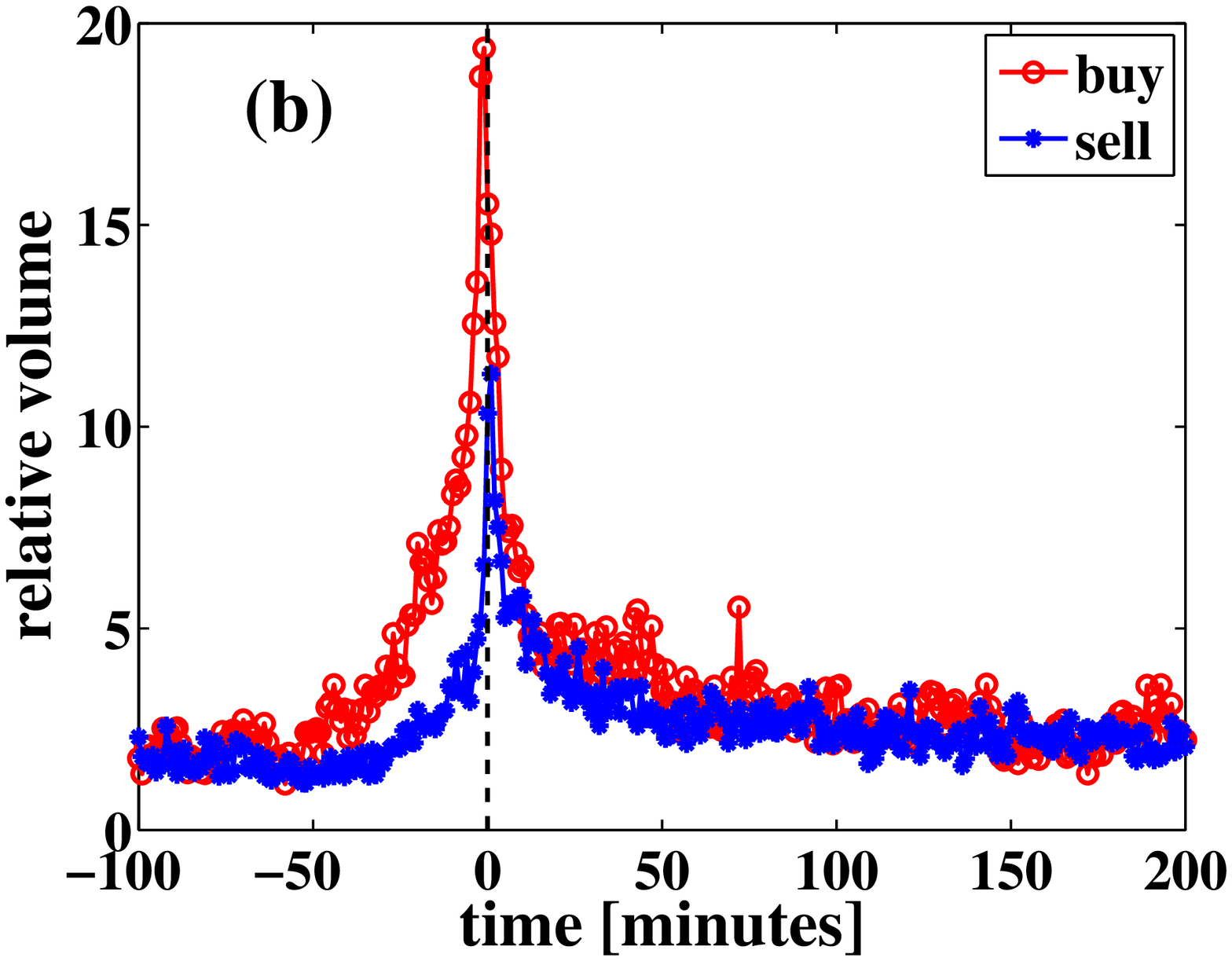}
\includegraphics[width=7cm]{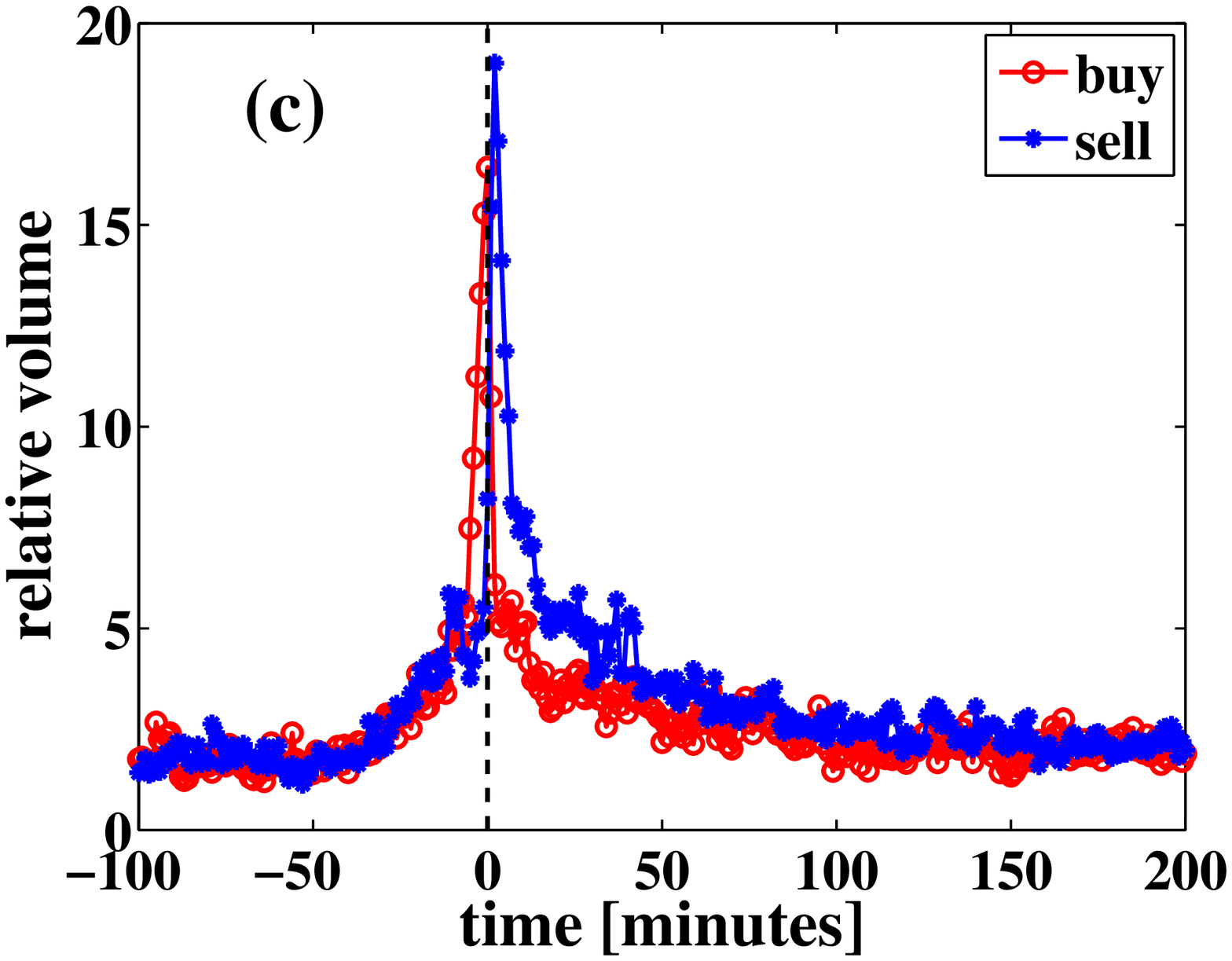}
\includegraphics[width=7cm]{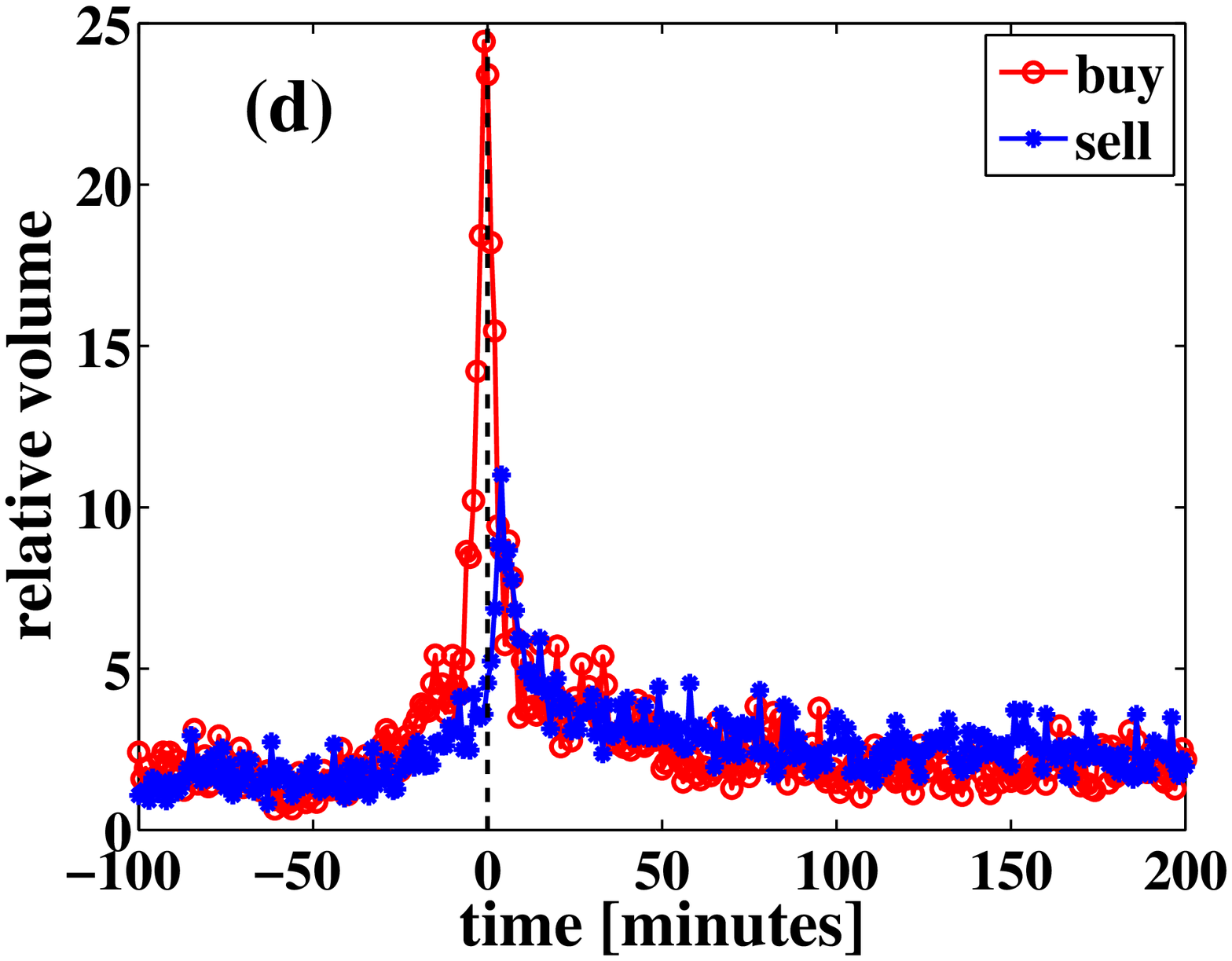}
\caption{\label{Fig:AggVol:Pos}  Dynamics of relative volume around 131 positive events for four types of buy orders and sell orders classified by order aggressiveness: (a) partially filled orders, (b) filled orders, (c) limit orders, and (d) canceled orders.}
\end{figure}

Figure \ref{Fig:AggVol:Pos} illustrates the dynamics of volumes around 131 positive events for the four types of buy orders and sell orders. Figure \ref{Fig:AggVol:Pos}a shows the results for partially filled market orders. The volume of buy orders that are partially filled starts to increase about 40 to 50 minutes before the extreme event and reaches about 22.4 times the normal value at $t_{\max}=0$. In contrast, the volume of sell orders that are partially filled market orders increases relatively slower and reaches its maximum of 10.3 at $t_{\max}=1$. Figure \ref{Fig:AggVol:Pos}b shows the results for filled market orders. The volume of buy filled orders starts to increase about 50 minutes before the extreme event and reaches about 19.4 times the normal value at $t_{\max}=-1$. In contrast, the volume of sell filled market orders increases relatively slower and reaches its maximum of 11.3 at $t_{\max}=1$. It is clear that sell market orders lag behind buy market orders and there are much more buy market orders before $t=0$ which push the price up. Figure \ref{Fig:AggVol:Pos}c shows the results for limit orders. The volume
of limit buy orders starts to increase about 50 minutes before the extreme event and reaches about 16.4 times the normal value at $t_{\max}=0$. In contrast, the volume of sell limit orders increases relatively slower and reaches its maximum of 19.0 at $t_{\max}=2$. Figure \ref{Fig:AggVol:Pos}d shows the results for canceled orders. We observe that the evolution behaviors are very similar to market orders in Fig.~\ref{Fig:AggVol:Pos}a and Fig.~\ref{Fig:AggVol:Pos}b. The maximum volume is 24.4 at $t_{\max}=-1$ for canceled buy orders and 19.0 at $t_{\max}=4$ for canceled sell orders. This reflects the fact that buy traders are very eager to execute their orders and more unfilled orders are canceled in order to submit new buy orders.

The characteristic values ($t_{\max}$ and $V_{\max}$) of the peaks in Fig.~\ref{Fig:AggVol:Pos} are listed in Table \ref{TB:AggVolNum:Pos}. The dynamics of numbers of the four order types around 131 positive events are very similar to that of volumes. However, there are still differences. The characteristic peak values ($t_{\max}'$ and $N_{\max}$) are also listed in Table \ref{TB:AggVolNum:Pos}. We find that $t_{\max}<t_{\max}'$ for filled buy orders and canceled orders and $t_{\max}=t_{\max}'$ for the rest six types of orders. After $t=-1$, the volume of filled buy order decreases while the number of filled buy orders still increases. It implies that investors (especially individuals) with small orders still rush in the market while investors (more probably institutions) has left the market.

\begin{table}[htb]
 \centering
 \caption{Peak characteristic values of the aggregate evolution trajectory of relative volume and relative number around 131 positive events for four types of buy orders and sell orders classified by order aggressiveness. The types of PFO, FO, LO and CO correspond to partially filled orders, filled orders, limit orders and canceled orders, respectively.}
 \label{TB:AggVolNum:Pos}
 \medskip
 \begin{tabular}{cccccccccccccccc}
  \hline\hline
   \multirow{3}*[1.5mm]{Type} && \multicolumn{4}{@{\extracolsep\fill}c}{Buy orders}  && \multicolumn{4}{@{\extracolsep\fill}c}{Sell orders}  \\
  \cline{3-6}\cline{8-11}
  && $t_{\max}$ & $V_{\max}$ & $t_{\max}'$ & $N_{\max}$ && $t_{\max}$ & $V_{\max}$ & $t_{\max}'$ & $N_{\max}$ \\\hline
  PFO &&  0 & 22.4 & 0 & 12.5 && 1 & 10.3 & 1 & 7.9 \\%
  FO && -1 & 19.4 & 1 & 12.9 && 1 & 11.3 & 1 & 12.2 \\%
  LO &&  0 & 16.4 & 0 & 11.7 && 2 & 19.0 & 2 & 16.9 \\%
  CO && -1 & 24.4 & 0 & 18.4 && 4 & 19.0 & 4 & 10.6\\%
  \hline\hline
 \end{tabular}
\end{table}

\begin{figure}[htb]
\centering
\includegraphics[width=3.5cm]{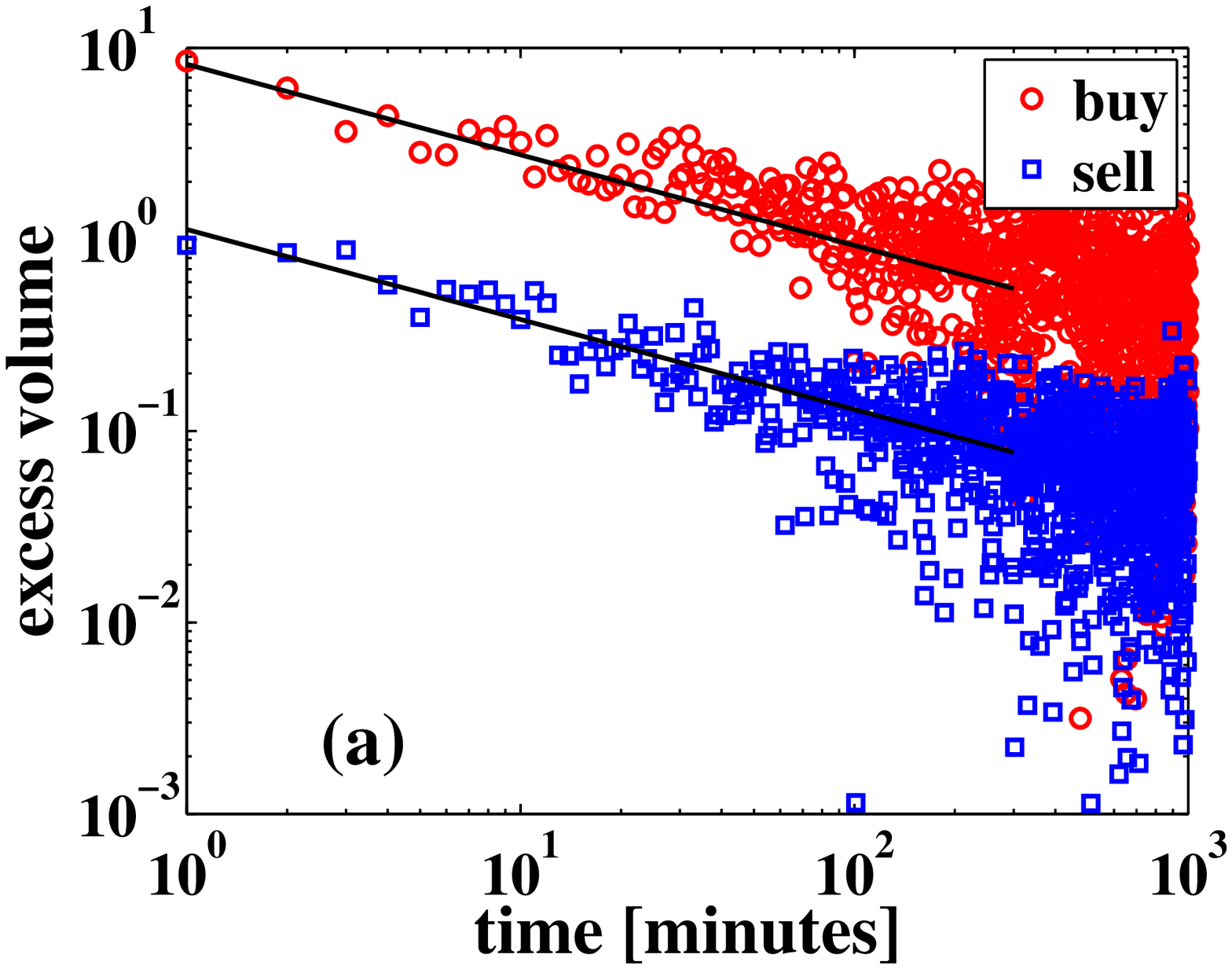}
\includegraphics[width=3.5cm]{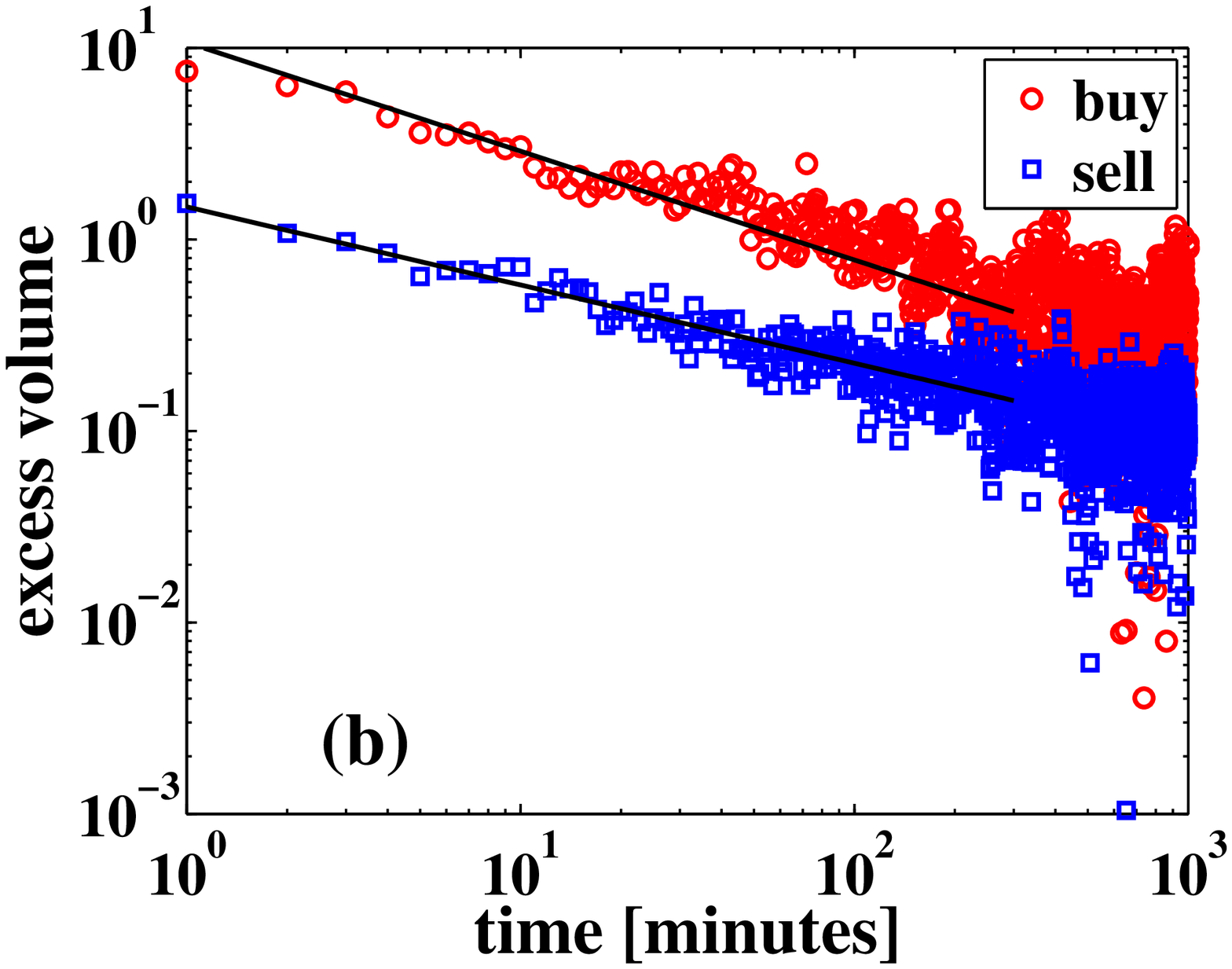}
\includegraphics[width=3.5cm]{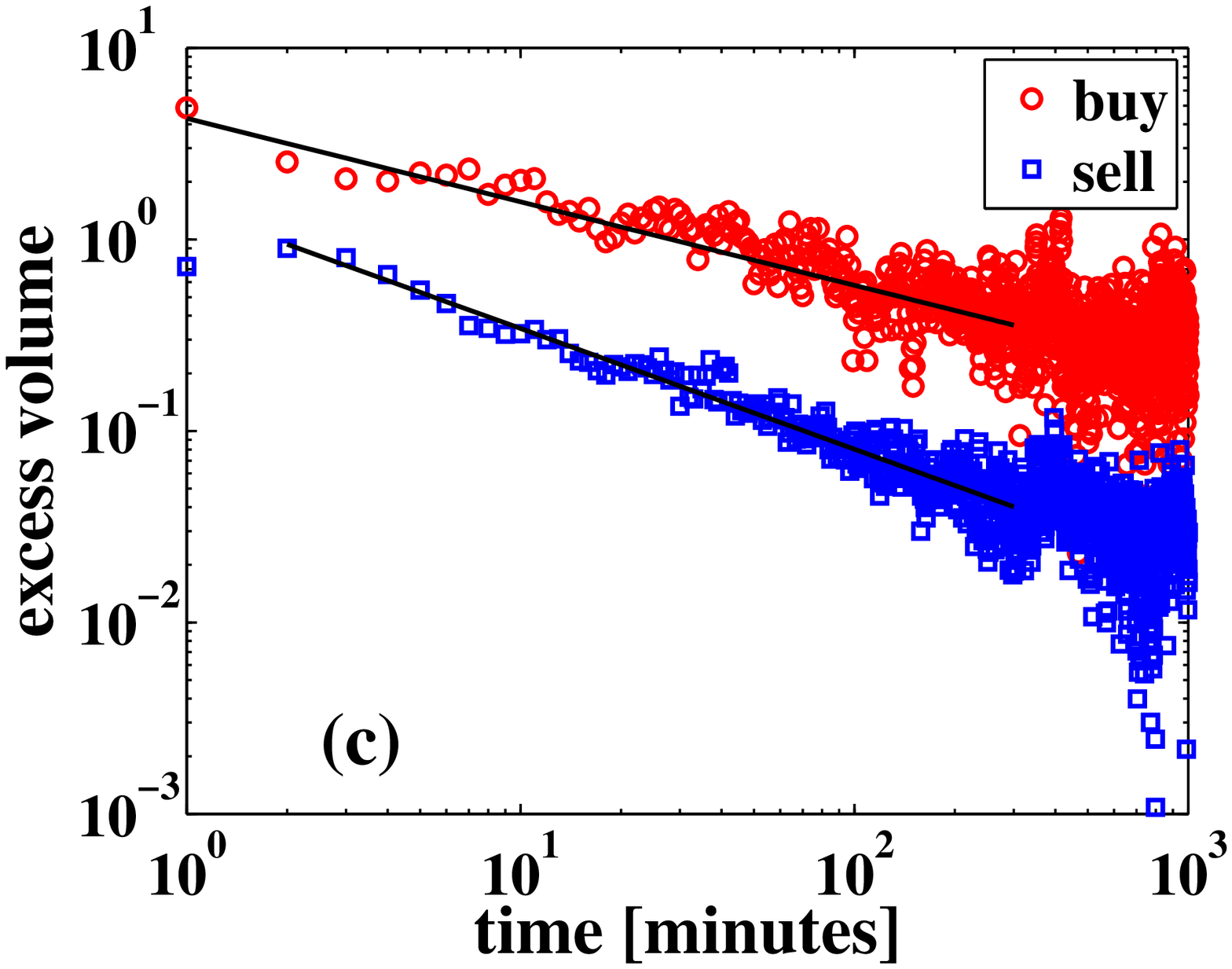}
\includegraphics[width=3.5cm]{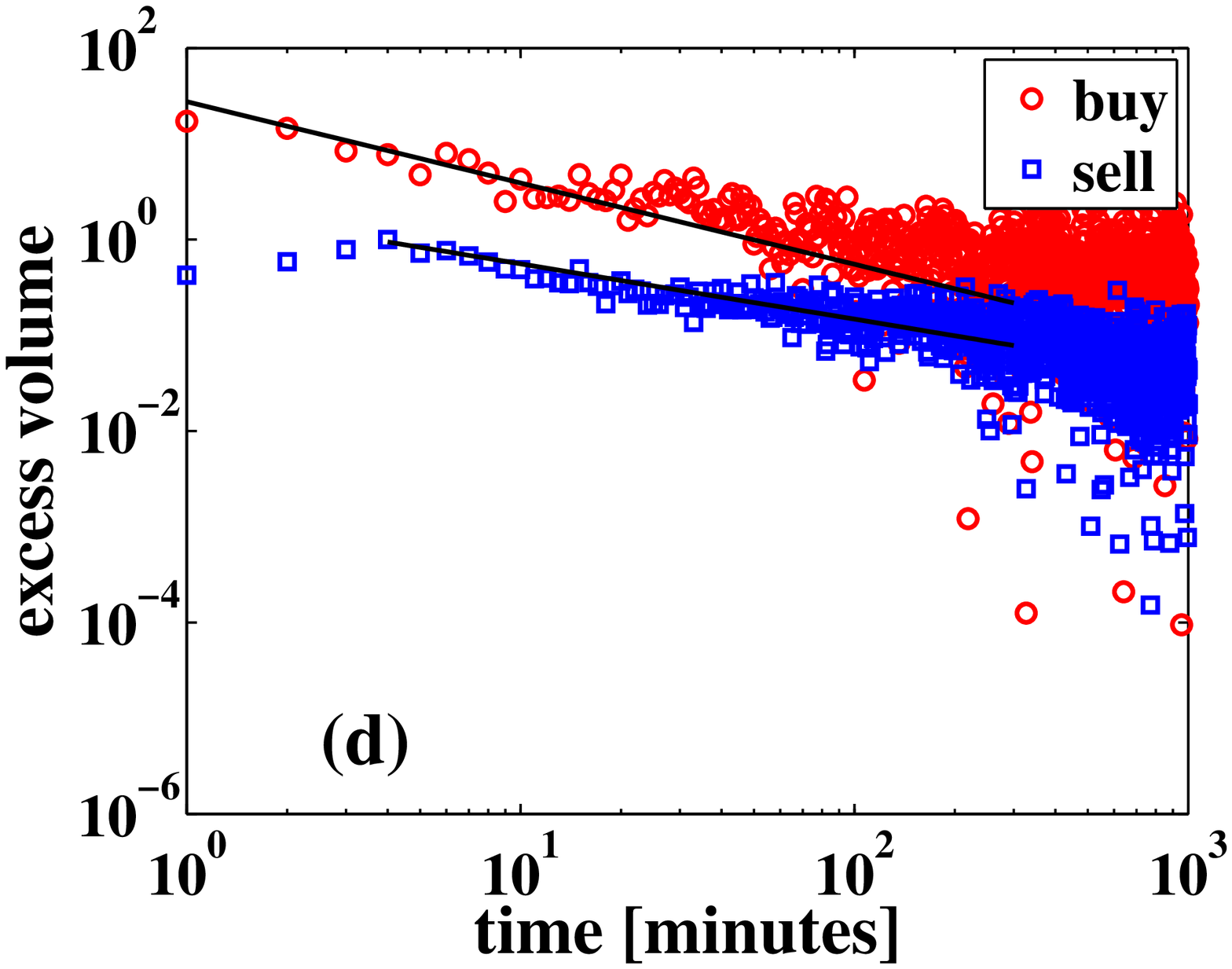}
\caption{\label{Fig:AggVol:Pos:PL}  Power-law
relaxation of excess volume around 131 positive events for four
types of buy orders and sell orders classified by order
aggressiveness: (a) partially filled orders, (b) filled orders, (c)
limit orders, and (d) canceled orders.}
\end{figure}

Figure \ref{Fig:AggVol:Pos:PL} shows the relaxation of excess volume after positive events ($t=0$) for the four types of buy and sell orders. All curves exhibit power-law behaviors. The power-law relaxation exponents are $0.47\pm0.05$ for partially filled buy orders in the scaling range $[1,300]$, $0.47\pm0.04$ for partially filled sell orders in the scaling range $[1,300]$, $0.57\pm0.04$ for filled buy orders in the scaling range $[1,300]$, $0.41\pm0.02$ for filled sell orders in the scaling range $[1,300]$, $0.43\pm0.03$ for buy limit orders in the scaling range $[1,300]$, $0.63\pm0.02$ for sell limit orders in the scaling range $[2,300]$, $0.85\pm0.09$ for canceled buy orders in the scaling range $[1,300]$, and $0.58\pm0.05$ for canceled sell orders in the scaling range $[4,300]$.

The excess order numbers after positive events ($t=0$) for the four types of buy and sell orders decay as power laws. The power-law relaxation exponents are $0.43\pm0.02$ for partially filled buy orders in the scaling range $[1,300]$, $0.46\pm0.02$ for partially filled sell orders in the scaling range $[1,300]$, $0.52\pm0.01$ for filled buy orders in the scaling range $[1,300]$, $0.44\pm0.01$ for filled sell orders in the scaling range $[1,300]$, $0.30\pm0.01$ for buy limit orders in the scaling range $[1,300]$, $0.65\pm0.01$ for sell limit orders in the scaling range $[2,300]$, $0.53\pm0.03$ for canceled buy orders in the scaling range $[1,300]$, and $0.54\pm0.02$ for canceled sell orders in the scaling range $[4,300]$.

\section{Individual and institutional investors}
\label{S1:Volume:Ind:Ins}

\subsection{Volume dynamics}

We investigate here whether the type of investors (individuals and institutions) has any effect on the volume dynamics of market orders around extreme price changes. Our data base contains the identifiers distinguishing the two types of investor. Specifically, we consider buy market orders and sell market orders by individuals and institutions around positive events and negative events (see Section \ref{S1:Buy:Sell}). In addition, we also consider partially filled buy and sell orders, filled buy and sell orders, buy and sell limit orders, and canceled buy and sell orders (see Section \ref{S1:Agg}). Since the findings are very similar, we present only the results for market buy and sell orders submitted by individuals and institutions around positive events.

\begin{figure}[htb]
\centering
\includegraphics[width=7cm]{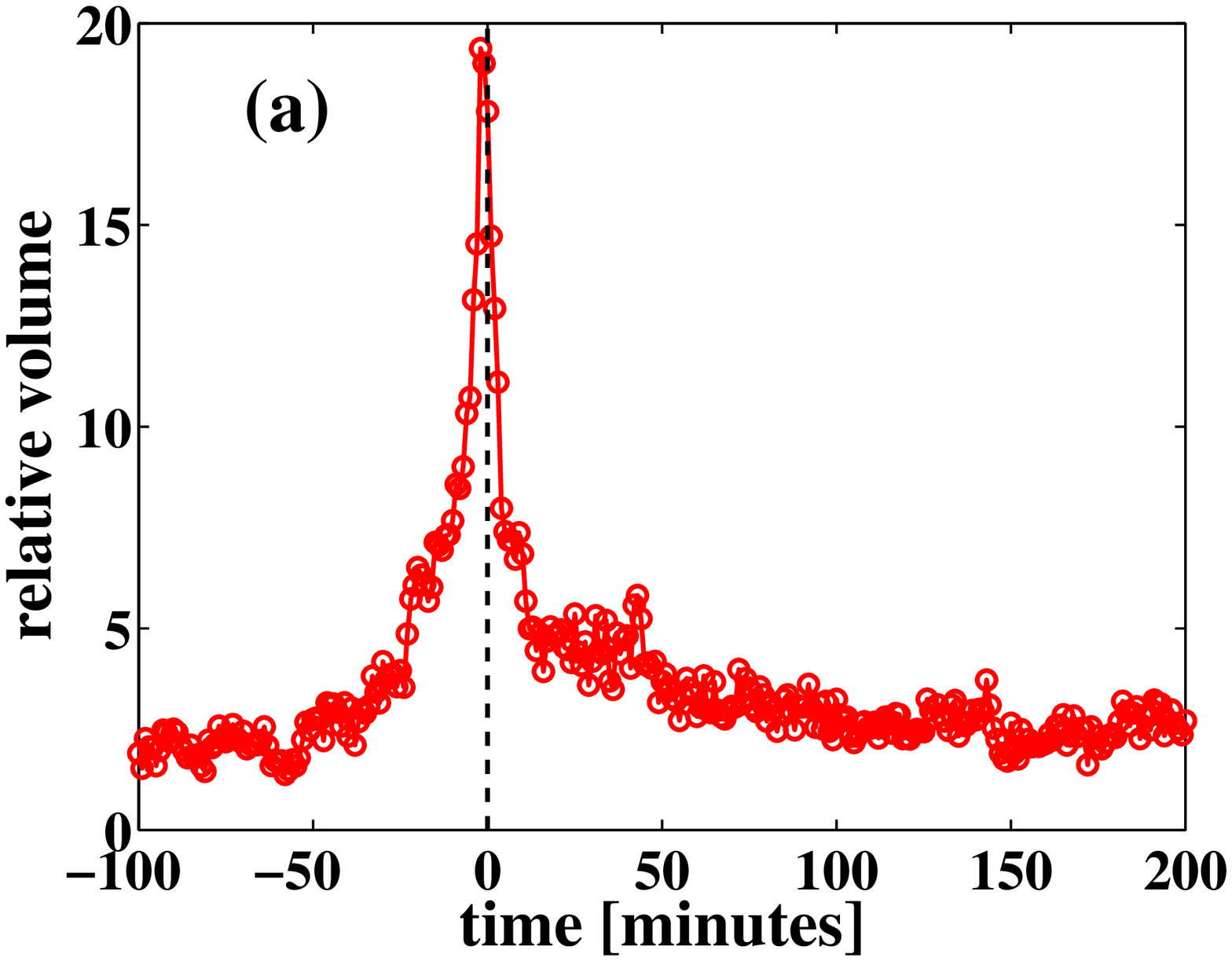}
\includegraphics[width=7cm]{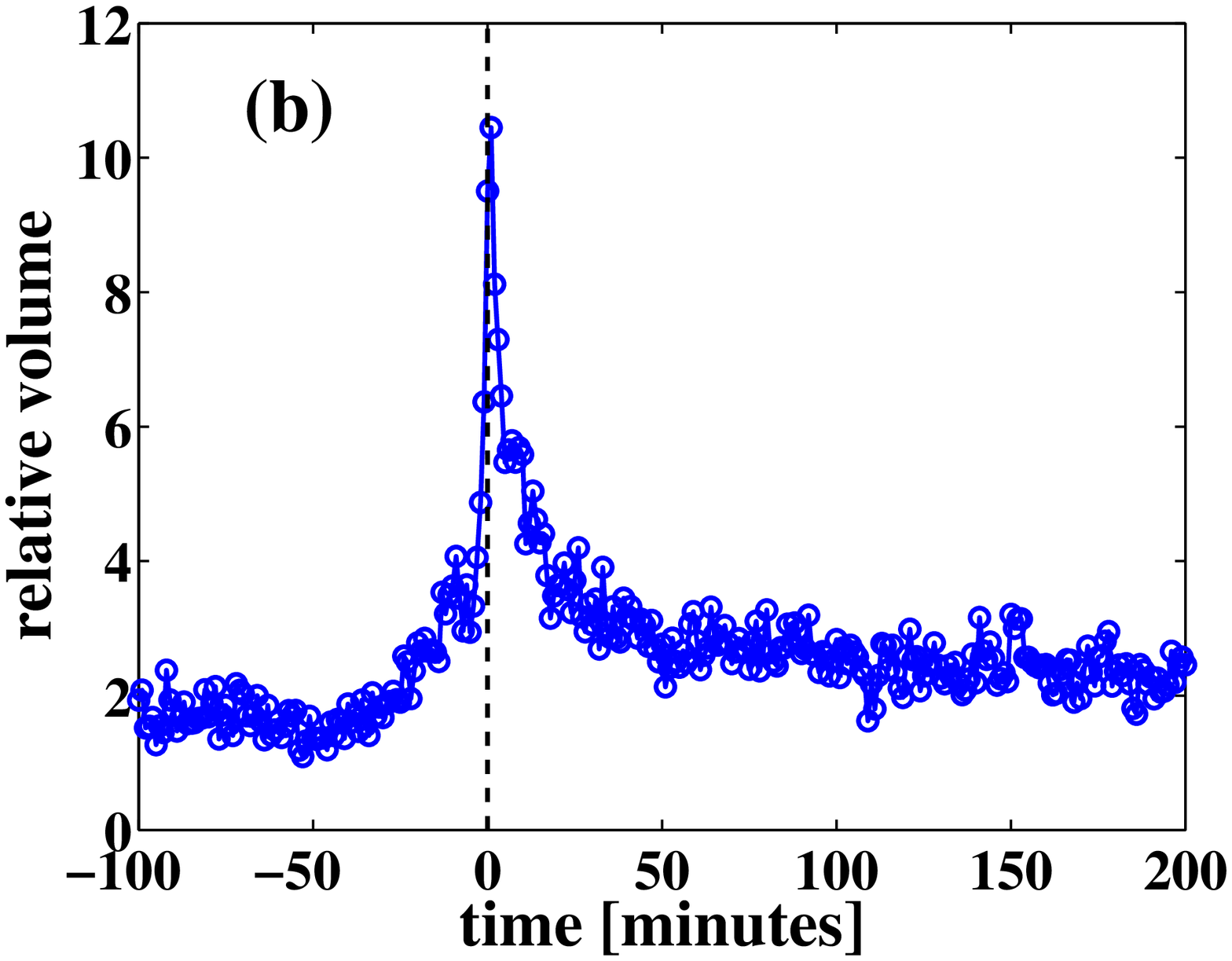}
\includegraphics[width=7cm]{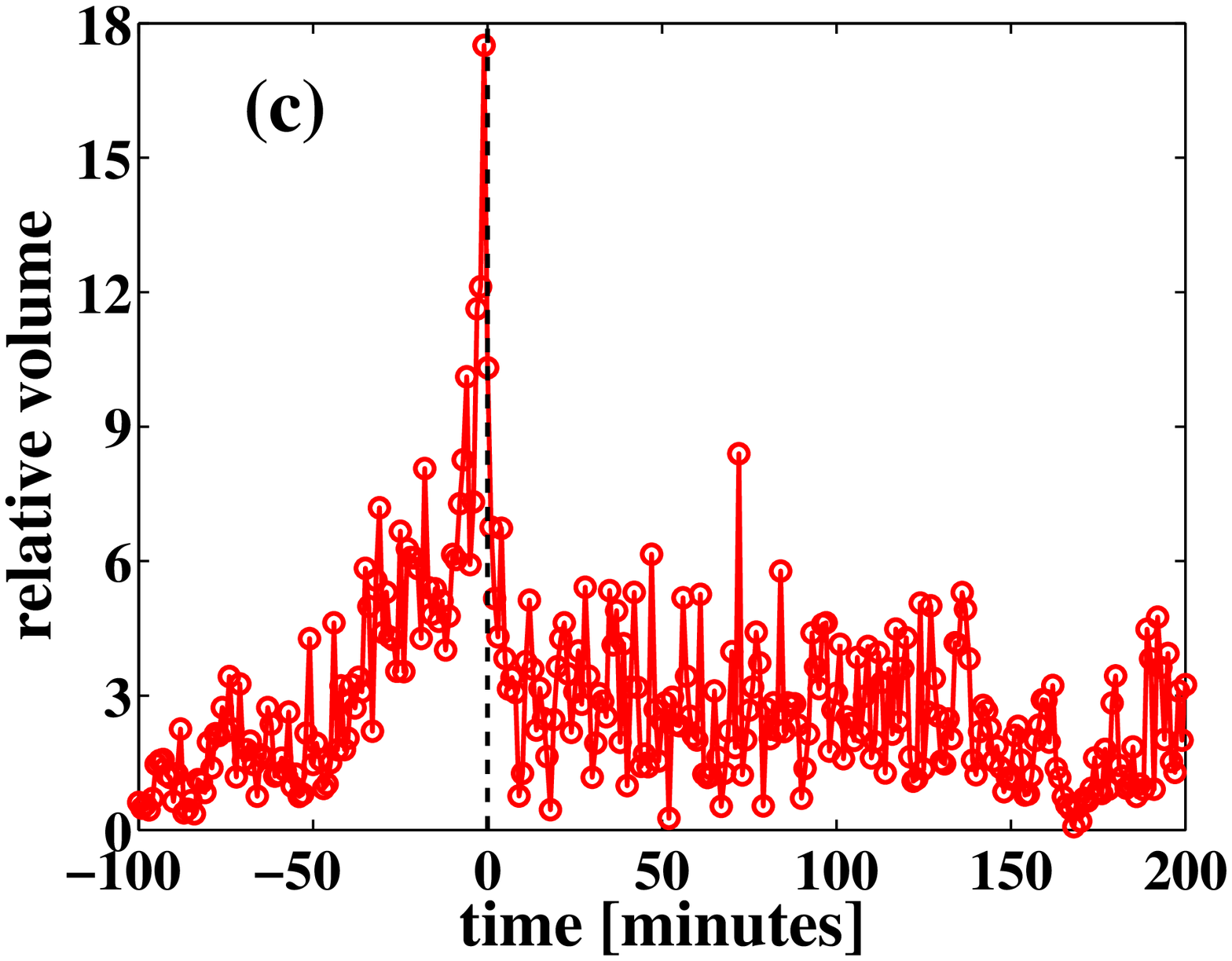}
\includegraphics[width=7cm]{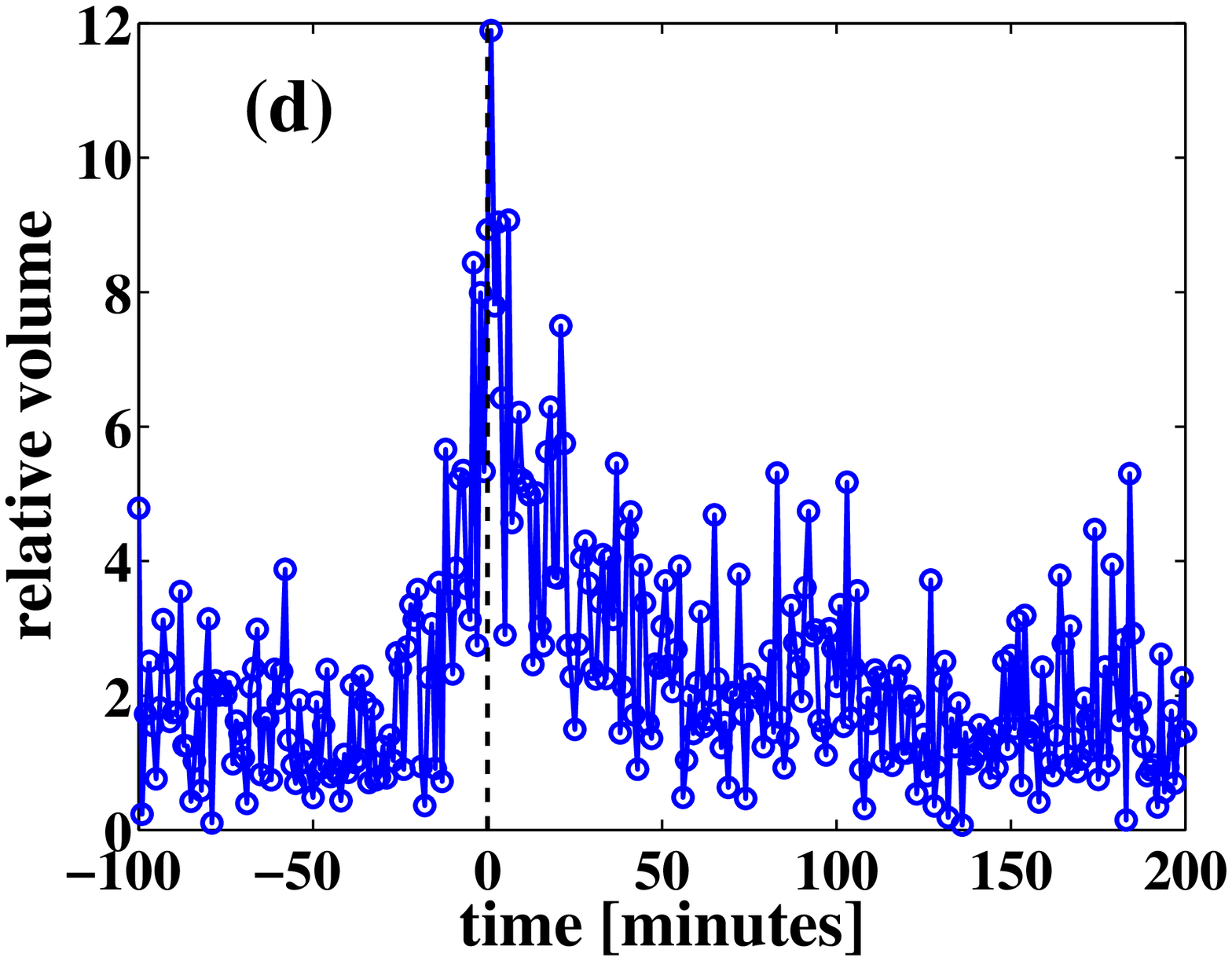}
\caption{\label{Fig:Volume:Ind:Ins}  Dynamics of relative volume of marketable orders submitted by individual and institutional investors around positive events: (a) marketable buy orders submitted by individuals, (b) marketable sell orders submitted by individuals, (c) marketable buy orders submitted by institutions, and (d) marketable sell orders submitted by institutions.}
\end{figure}

The results are shown in Fig.~\ref{Fig:Volume:Ind:Ins}. Comparing the volume dynamics of buy (or sell) market orders from individuals and institutions, it seems that the type of investors has negligible effect on the dynamics. The volume dynamics of buy (resp. sell) market orders from individuals and institutions are very similar to that shown in Fig.~\ref{Fig:Direction}a (resp. Fig.~\ref{Fig:Direction}b). We observe that $t_{\max}=-2$ and $V_{\max}=19.4$ for buy market orders submitted by individuals, $t_{\max}=-1$ and $V_{\max}=17.5$ for buy market orders submitted by institutions, $t_{\max}=1$ and $V_{\max}=10.4$ for sell market orders submitted by individuals, $t_{\max}=1$ and $V_{\max}=11.9$ for buy market orders submitted by institutions. The values of $t_{\max}$ for individuals and institutions are identical to that shown in Fig.~\ref{Fig:Direction}a for buy market orders and Fig.~\ref{Fig:Direction}b for sell market orders respectively except for buy market orders from individuals.

\subsection{Dynamics of relative rates of market orders, limit orders and cancelations}
\label{S2:Rate:Aggr}

\begin{figure}[htb]
\centering
\includegraphics[width=7cm]{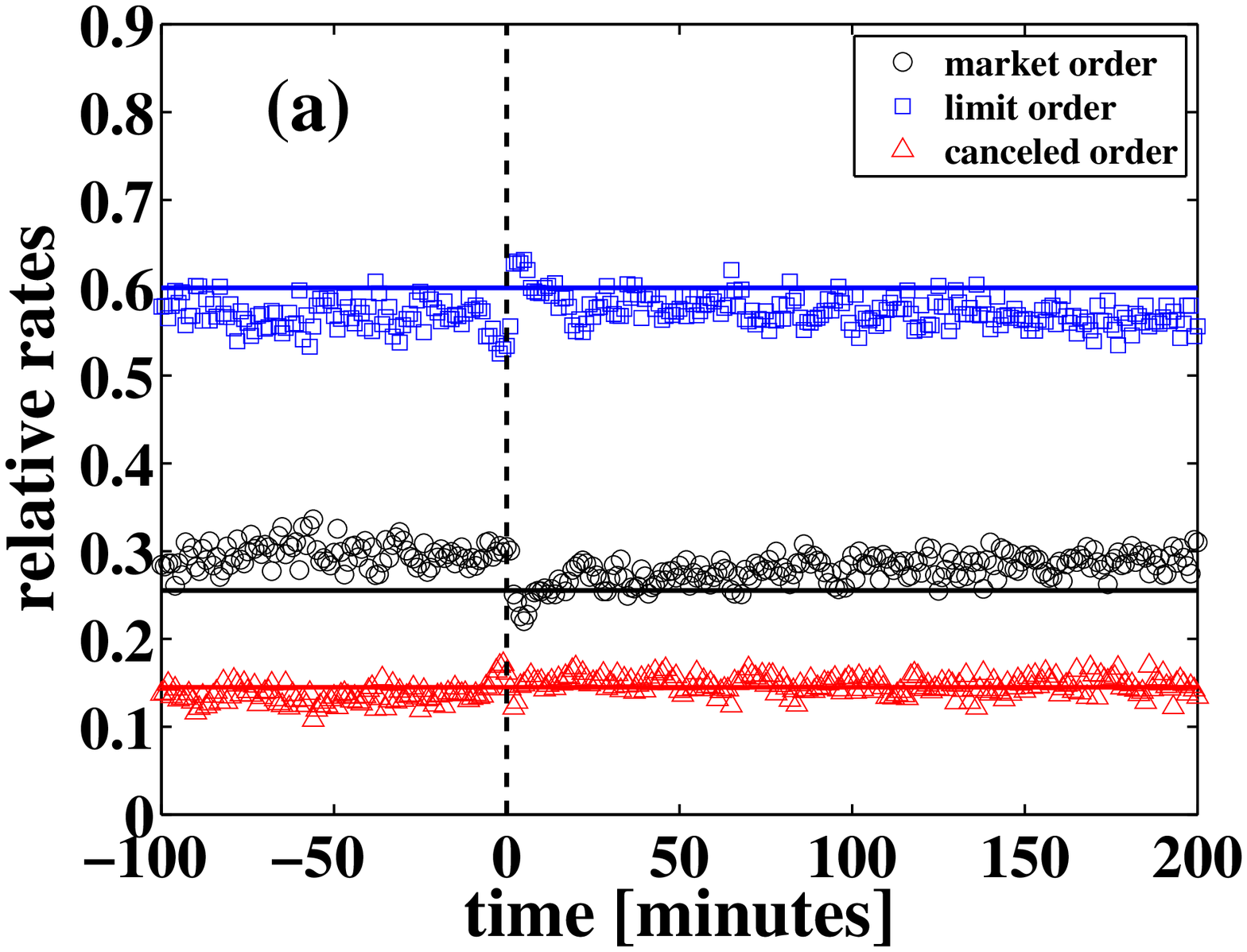}
\includegraphics[width=7cm]{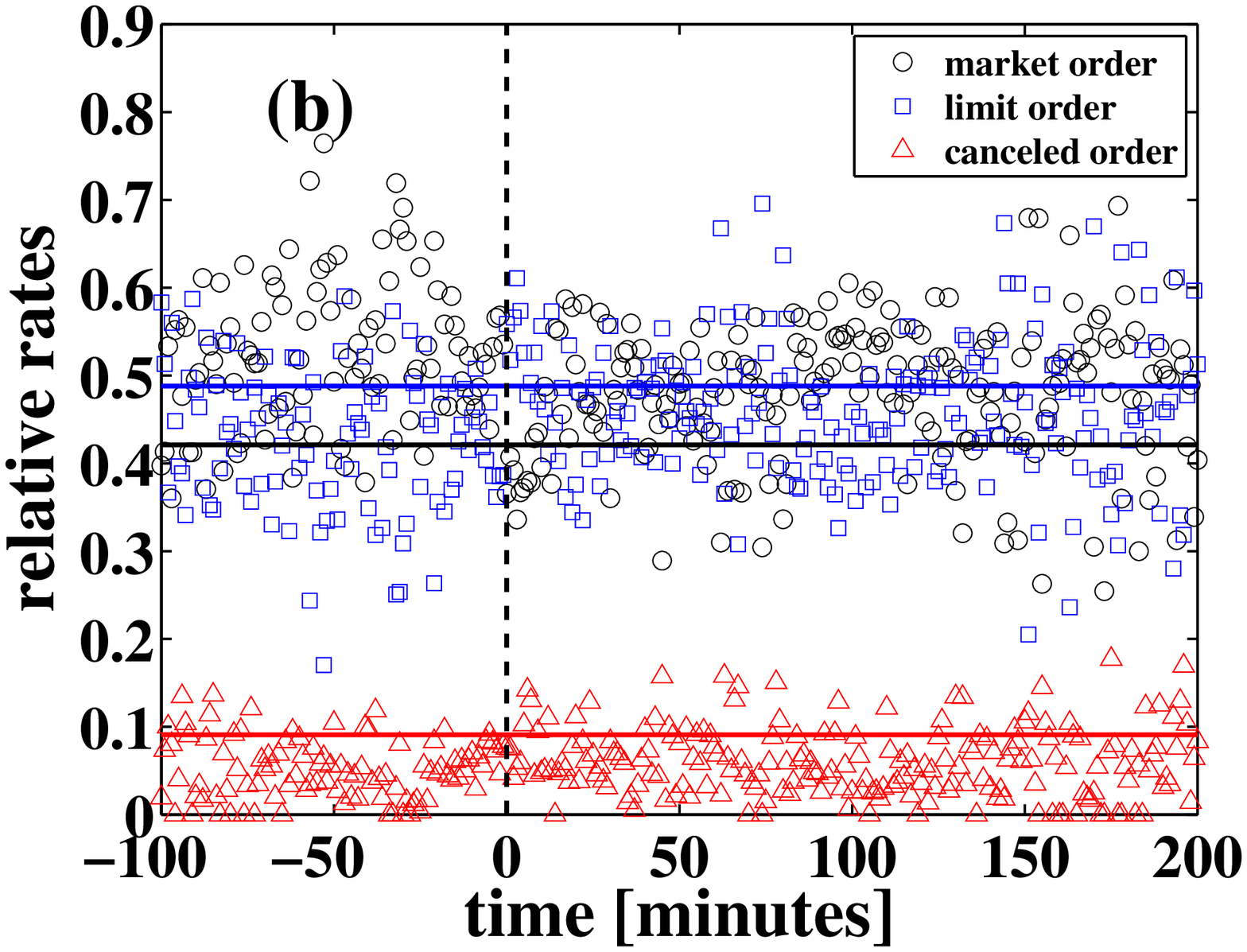}
\caption{\label{Fig:Rate:Ind:Ins}  Dynamics of the relative rates of market orders (circles ${\circ}$), limit orders (squares $\color{blue}{\square}$) and cancelations (triangles $\color{red}{\triangle}$) submitted by (a) individuals and (b) institutions around positive events. The horizontal lines close to the symbols are the corresponding average rates of the three types of orders in normal periods.}
\end{figure}

Following Ref.~\cite{Toth-Kertesz-Farmer-2009-EPJB}, we study in this section the dynamics of relative rates of different types of orders (market orders, limit orders and canceled orders) around extreme events. Furthermore, we distinguish each type of orders by the type of investors. For each type of investors, the numbers of market orders, limit orders and cancelations submitted in the same minute are determined and the relative order rates are calculated as the proportions of order numbers. Figure \ref{Fig:Rate:Ind:Ins} shows the results for positive events. We do not show the results for negative events since the conclusion is the same. We find no very strong variations in the different rates around the large price changes. This finding is qualitatively the same as the LSE stocks \cite{Toth-Kertesz-Farmer-2009-EPJB}. It seems, however, justified to say that the values of the relative rates are not identical between SZSE stocks and LSE stocks.

In Fig.~\ref{Fig:Rate:Ind:Ins}, the relative rates around extreme events are compared with those in normal periods. For individuals, the average relative rates in normal periods for market orders, limit orders, and canceled orders are 0.26, 0.60, and 0.14, respectively. Before the extreme events at $t=0$, the relative rate of market orders is higher than the normal level, while the relative rates of limit orders and canceled orders are lower than the normal level. After the extreme events, the situation is different. The relative rate of limit orders jumps above the normal level right after the events and then decreases to its previous level before the events in about 15 minutes. In contrast, the relative rate of market orders drops under its normal level immediately after the events and then returns to its previous level before the events. The relative rate of canceled orders becomes identical to the normal level. For institutions, the average relative rates in normal periods for market orders, limit orders, and canceled orders are 0.42, 0.49, and 0.09, respectively. On average, the relative rate of market orders is higher than the normal level, while the those for the limit orders and canceled orders are lower than the corresponding normal levels. The abrupt change right after the events are visible but not significant since the data are very noisy.

There are differences between the behavior of relative rates for SZSE stocks and LSE stocks. For 12 LSE stocks, there are no strong changes in the rates for {\em{true}} market orders, {\em{true}} limit orders and canceled orders around large price changes, and their rates fluctuate around $\sim 0.07$, $\sim0.61$ and $\sim 0.32$ \cite{Toth-Kertesz-Farmer-2009-EPJB}. This result means that no strategy of the investors about how they place orders is identified \cite{Toth-Kertesz-Farmer-2009-EPJB}. After accounting for the effective orders, the relative rates of limit orders, market orders, and cancelations for 14 LSE stocks are 0.16, 0.49 and 0.35 \cite{Farmer-Gillemot-Lillo-Mike-Sen-2004-QF}. We see that the cancelation ratio of the LSE stocks is much higher than that of the SZSE stocks, while the rate of market order is much lower than that of the SZSE stocks. If we consider only the orders submitted, the ratio of the number of effective market orders to the total number of limit and market orders are 0.30 and 0.25 for the SZSE stocks and the LSE stocks, respectively. This observation can be attribute to the facts that the proportion of institutional investors is much higher in the LSE and that the LSE is a mature market while the SZSE is an emerging market.

Comparing the results for individual and institutions, we see significant differences. For individuals, the relative rate of limit orders is about 0.58, much larger than that of market orders (about 0.30). In contrast, for institutions, the relative rate of market order (about 0.50) is larger than that of limit orders (about 0.45). In addition, institutions have a smaller rate of canceled orders than individuals. The difference between the behavior of the individual and institutional investors is very interesting since it sheds light onto the strategies of these groups. It may indicate that institutions are more aggressive than individuals thus institutions play a more important role in causing large price changes. Concerning the relative rate of market orders after the positive event, a careful scrutiny unveils an increasing trend for individuals and a noisy decreasing trend for institutions. It means that individuals are still aggressively submit buy market orders after the peak when the institutions are quitting gradually. Combining together the fact that the relative rate of market orders is much higher for institutional investors, we can draw a conclusion that institutions are more informed than individuals in the Chinese stock markets. This is consistent with the conventional wisdom and we can conjecture that individuals have worse performs than institutions \cite{Barber-Lee-Liu-Odean-2008-RFS}.

\subsection{Dynamics of relative rates of buy and sell market orders, limit orders and cancelations}
\label{S2:Rate:Aggr:Sign}

\begin{figure}[htb]
\centering
\includegraphics[width=7.5cm]{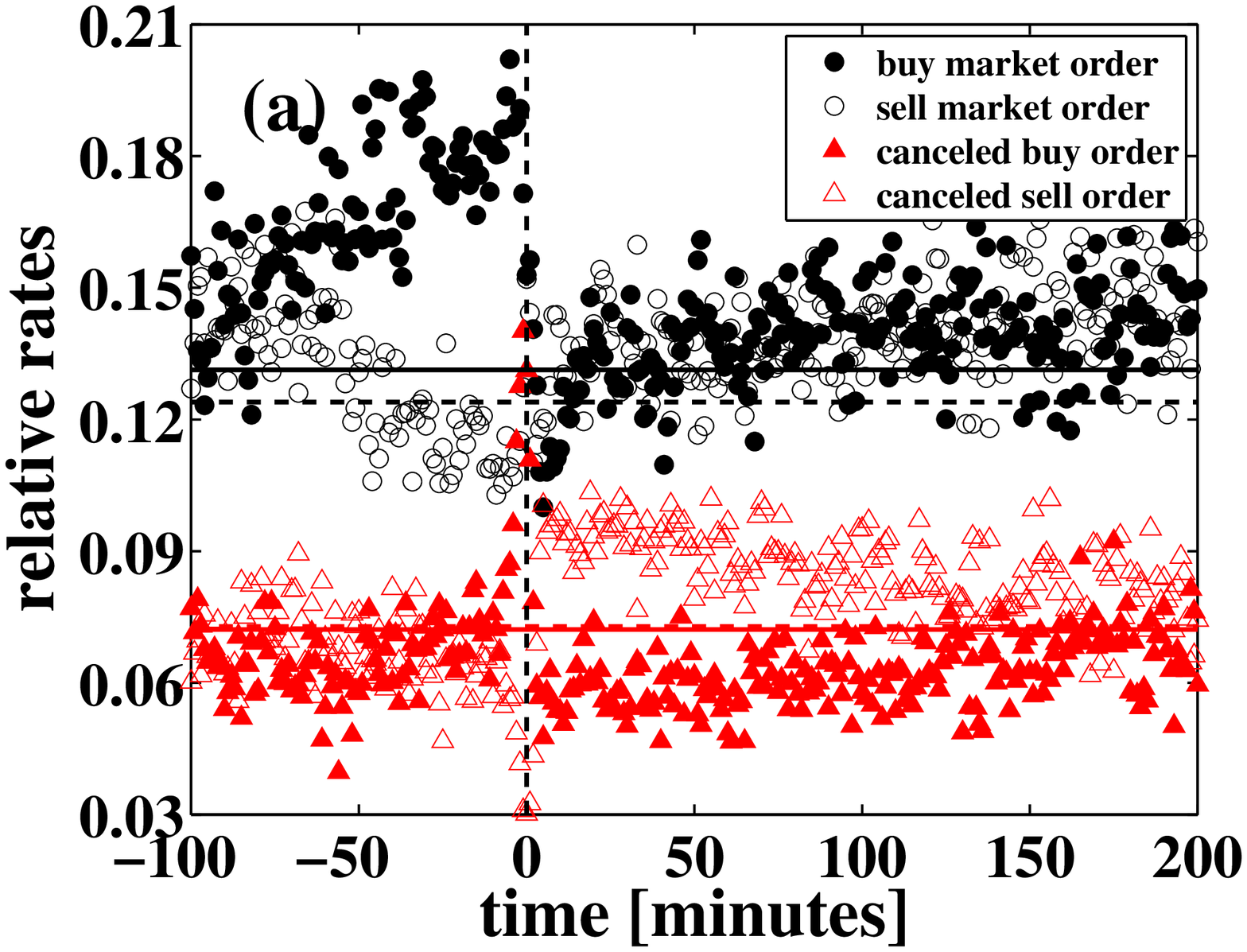}
\includegraphics[width=7.5cm]{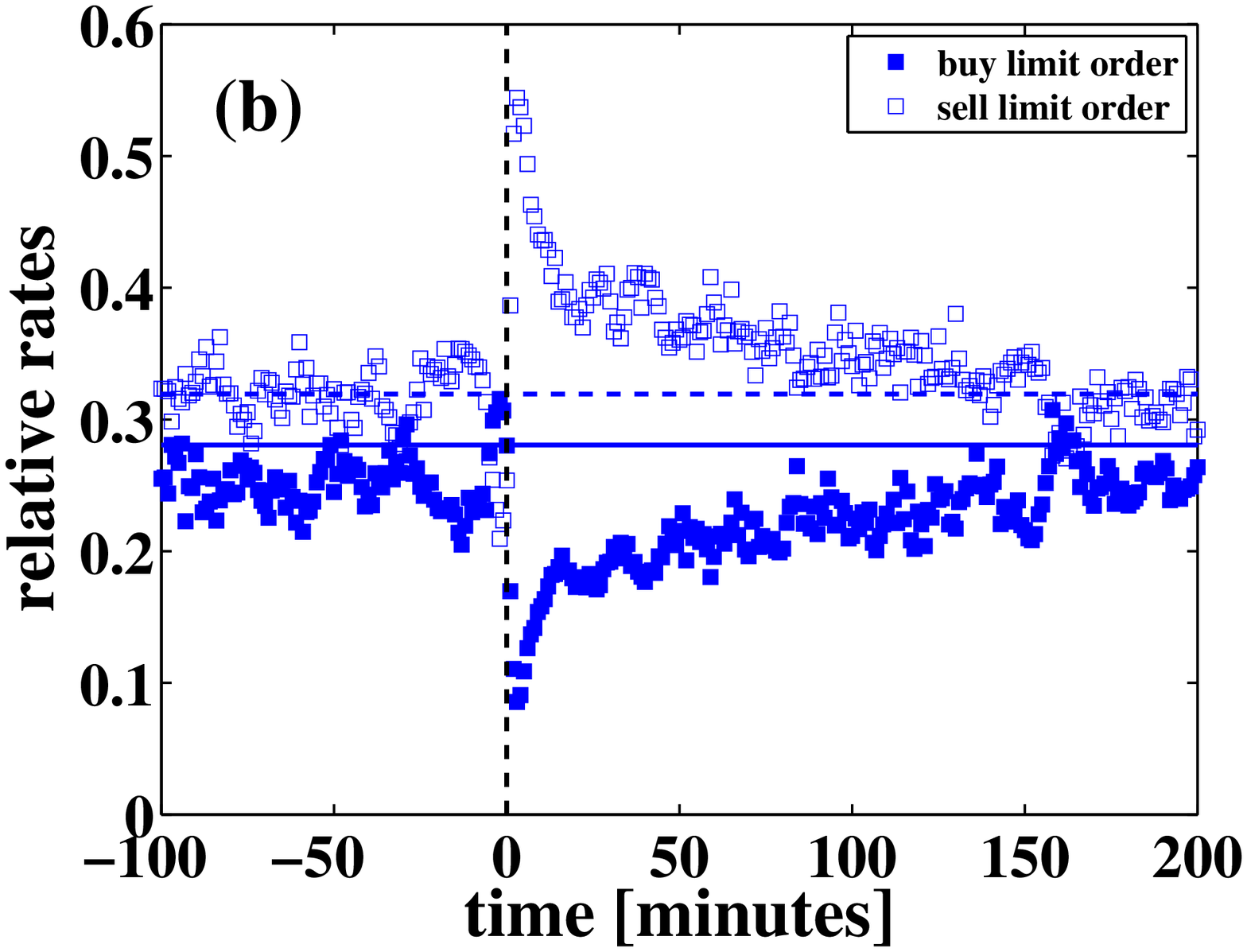}
\includegraphics[width=7.5cm]{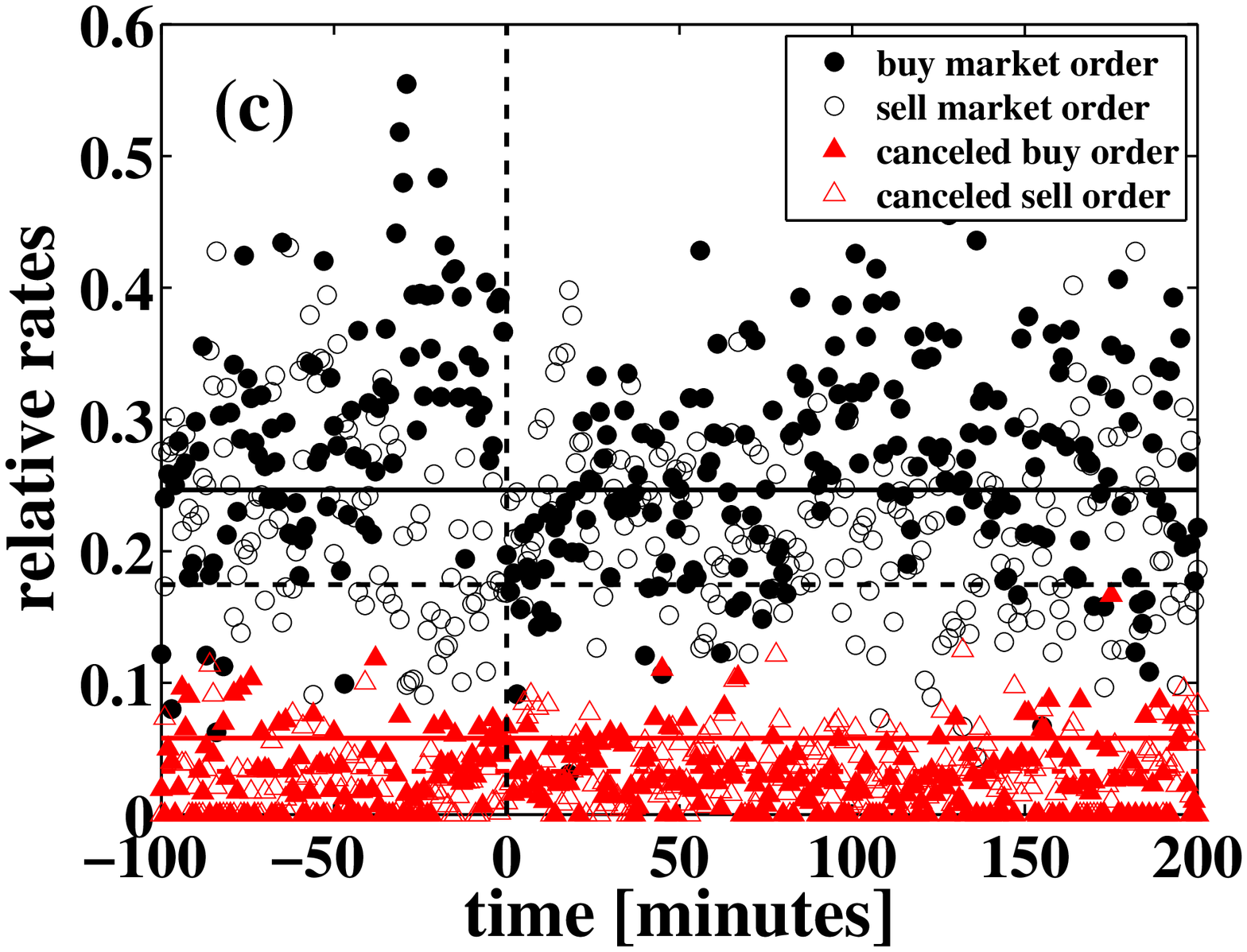}
\includegraphics[width=7.5cm]{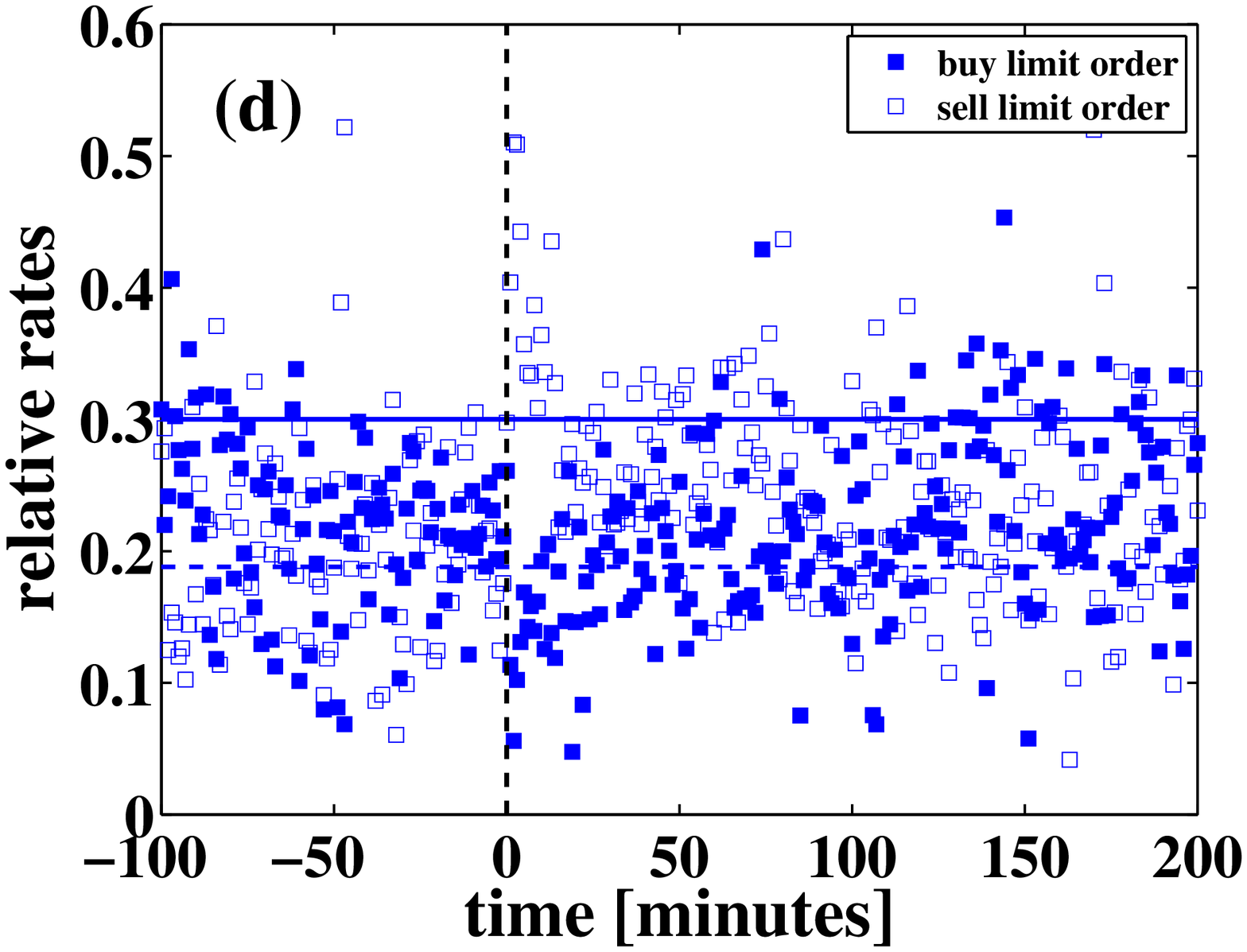}
\caption{\label{Fig:Rate:Ind:Ins:Buy:Sell}  Dynamics of the relative rates of market orders (circles ${\circ}$), limit orders (squares $\color{blue}{\square}$) and cancelations (triangles $\color{red}{\triangle}$) around positive events. Solid symbols denote buy orders, while open symbols denote sell orders. (a) Relative rates of market orders and canceled orders submitted by individuals. The average relative rates in normal periods are 0.13 for buy market orders, 0.12 for sell market orders, 0.07 for canceled buy orders, and 0.07 for canceled sell orders.  (b) Relative rate of limit orders submitted by individuals. The average relative rates in normal periods are 0.28 and 0.32 for buy and sell limit orders. (c) Relative rate of market orders and canceled orders submitted by institutions. The average relative rates in normal periods are 0.25 for buy market orders, 0.17 for sell market orders, 0.06 for canceled buy orders, and 0.03 for canceled sell orders. (d) Relative rate of limit orders submitted by institutions. The average relative rates in normal periods are 0.30 and 0.19 for buy and sell limit orders.}
\end{figure}

The difference between the behaviors of individuals and institutions is more evident when we further distinguish buy orders and sell orders. Figure \ref{Fig:Rate:Ind:Ins:Buy:Sell}a and Fig.~\ref{Fig:Rate:Ind:Ins:Buy:Sell}b illustrate the results for individuals, while Fig.~\ref{Fig:Rate:Ind:Ins:Buy:Sell}c and Fig.~\ref{Fig:Rate:Ind:Ins:Buy:Sell}d show the results for institutions. We find a strong buy-sell asymmetry in both types of investors. According to Fig.~\ref{Fig:Rate:Ind:Ins:Buy:Sell}a, the relative rate of buy market orders submitted by individuals increases continuously before the extreme events and then drops sharply below its normal level with a fast recovery above its normal level, while the relative rate of sell market orders decreases continuously before the extreme events and recovers soon to its normal level smoothly without a sharp increase above its normal level. The relative rate of canceled buy orders also increases to reach a high level before the extreme events and then drops rapidly to its normal level, while the relative rate of canceled sell orders decreases before the extreme events and increase suddenly to a high value which is followed by a slow relaxation to its normal level. According to Figure \ref{Fig:Rate:Ind:Ins:Buy:Sell}b, the two curves of the relative rates of buy limit orders and sell limit orders are almost symmetric to the mean of their normal values. The rate of buy limit orders increases before the extreme events and decreases markedly, which is followed by a slow recovery to its normal level.

It turns out that for institutional investors the relative rates of the buy and sell orders behave in an opposite way such that the difference roughly cancels. This can be observed for both market and limit orders. On the other hand, though the results are very noisy, the difference seems somewhat smaller for market orders and significantly smaller for limit orders. It seems as the individual investors would simply follow the events, while the institutional ones act according to some strategies. We also find that there are less buy limit orders canceled by institutions and there are no cancelation of buy orders after all in many minutes around positive events. All these point towards the stronger aggressiveness of the institutional investors.

\section{Conclusion}
\label{S1:conclusion}

We studied the dynamics of order flows around large price changes in 23 Chinese stocks traded on the Shenzhen Stock Exchange in 2003. Special emphasis was put on the dynamics of order volume. We found sound evidence of price reversal after large price changes. In the case of positive event, the price soars up about 5\% in 100 minutes and then drops about 1\% in 15 minutes to a constant level. This reversal pattern indicates that the price impact of orders around positive events is permanent. The situation around negative events is similar except that the price after large negative events does not seem to relax to a constant level. It should be noted that due to the much lower number of negative events the results for them are always much more noisy than for the positive ones.

For the dynamics of volatility (absolute return), volume, bid-ask spread, and volume imbalance, we found a significant peak around the extreme price changes followed by a slow relaxation, which can be well fitted by a power law. In most cases, the relaxation exponents are larger for the SZSE stocks than for the LSE stocks.

We investigated the effect of order direction, order aggressiveness and investor type on the dynamics of order flows. In each case, we found an increase of order volume followed by a slow decrease, which forms a peak around the extreme events. The volume peak of buyer initiated execution orders was found to appear earlier and higher than seller initiated executed orders around positive events, while the volume peak of seller initiated execution orders was found to lead buyer initiated execution orders in the magnitude and time around negative events. When orders are divided into four groups according to their aggressiveness, the behaviors of partially filled market orders, filled market orders and canceled orders are qualitatively the same: The volume peak of buy (sell) orders of each type appears earlier and higher than sell (buy) orders around positive (negative) events. However, the volume peak of sell (buy) limit orders shows up earlier and higher than buy (sell) orders around positive (negative) events. No qualitative difference has been observed in the volume dynamics around extreme events between individuals and institutions.

We also studied the relative rates of different types of orders. We confirmed that the relative rates of market orders, limit orders and cancelations do not fluctuate much around extreme events, but there is still a significant difference between individuals and institutions. In addition, we witnessed differences in the dynamics around extreme events for buy orders and sell orders. Interestingly, the behaviors are different when the type of investors is taken into consideration. Clearly, there is a sharp difference between the strategies of individual and institutional investors, and institutions are more aggressive and more informed.

\section*{Acknowledgments:}

GHM and WXZ acknowledge support from the ``Shu Guang'' project sponsored by Shanghai Municipal Education Commission and Shanghai Education Development Foundation (2008SG29) and the Program for New Century Excellent Talents in University (NCET-07-0288). JK acknowledges support from COST MP0801 as well as from OTKA K60456 and T049238.

\bibliographystyle{iopart-num} 
\bibliography{E:/Papers/Auxiliary/Bibliography}

\providecommand{\newblock}{}
\begin{thebibliography}{10}
\expandafter\ifx\csname url\endcsname\relax
  \def\url#1{{\tt #1}}\fi
\expandafter\ifx\csname urlprefix\endcsname\relax\def\urlprefix{URL }\fi
\providecommand{\eprint}[2][]{\url{#2}}

\bibitem{Sornette-Woordard-2009-XXX}
Sornette D and Woodard R 2009 {Financial bubbles, real estate bubbles,
  derivative bubbles, and the financial and economic crisis} arXiv: 0905.0220

\bibitem{Sornette-Zhou-2004-PA}
Sornette D and Zhou W~X 2004 {\em Physica A\/} {\bf 332} 412--440

\bibitem{Johansen-Sornette-2000a-EPJB}
Johansen A and Sornette D 2000 {\em Eur. Phys. J. B\/} {\bf 17} 319--328

\bibitem{Zhou-Sornette-2004b-PA}
Zhou W~X and Sornette D 2004 {\em Physica A\/} {\bf 337} 586--608

\bibitem{Zhou-Sornette-2006b-PA}
Zhou W~X and Sornette D 2006 {\em Physica A\/} {\bf 361} 297--308

\bibitem{Sornette-Woordard-Zhou-2009-PA}
Sornette D, Woodard R and Zhou W~X 2009 {\em Physica A\/} {\bf 388} 1571--1576

\bibitem{Sornette-2003-PR}
Sornette D 2003 {\em Phys. Rep.\/} {\bf 378} 1--98

\bibitem{Cont-2001-QF}
Cont R 2001 {\em Quant. Financ.\/} {\bf 1} 223--236

\bibitem{Johnson-Jefferies-Hui-2003}
Johnson N~F, Jefferies P and Hui P~M 2003 {\em {Financial Market Complexity:
  What Physics Can Tell Us About Market Behaviour}\/} (New York: Oxford
  University Press)

\bibitem{Lillo-Mantegna-2004-PA}
Lillo F and Mantegna R~N 2004 {\em Physica A\/} {\bf 338} 125--134

\bibitem{Selcuk-2004-PA}
Sel{\c{c}}uk F 2004 {\em Physica A\/} {\bf 333} 306--316

\bibitem{Selcuk-Gencay-2006-PA}
Sel{\c{c}}uk F and Gen{\c{c}}ay R 2006 {\em Physica A\/} {\bf 367} 375--387

\bibitem{Weber-Wang-VodenskaChitkushev-Havlin-Stanley-2007-PRE}
Weber P, Wang F~Z, Vodenska-Chitkushev I, Havlin S and Stanley H~E 2007 {\em
  Phys. Rev. E\/} {\bf 76} 016109

\bibitem{Mu-Zhou-2008-PA}
Mu G~H and Zhou W~X 2008 {\em Physica A\/} {\bf 387} 5211--5218

\bibitem{Jiang-Li-Cai-Wang-2009-PA}
Jiang J, Li W, Cai X and Wang Q~P~A 2009 {\em Physica A\/} {\bf 388} 1893--1907

\bibitem{Sornette-Johansen-Bouchaud-1996-JPIF}
Sornette D, Johansen A and Bouchaud J~P 1996 {\em J. Phys. I France\/} {\bf 6}
  167--175

\bibitem{Zawadowski-Kertesz-Andor-2004-PA}
Zawadowski A~G, Kert{\'e}sz J and Ador G 2004 {\em Physica A\/} {\bf 344}
  221--226

\bibitem{Zawadowski-Andor-Kertesz-2006-QF}
Zawadowski A~G, Ador G and Kert{\'e}sz J 2006 {\em Quant. Financ.\/} {\bf 6}
  283--295

\bibitem{Toth-Kertesz-Farmer-2009-EPJB}
T{\'o}th B, Kert{\'e}sz J and Farmer J~D 2009 {\em Eur. Phys. J. B\/} {\bf 71}
  499--510

\bibitem{Ponzi-Lillo-Mantegna-2009-PRE}
Ponzi A, Lillo F and Mantegna R~N 2009 {\em Phys. Rev. E\/} {\bf 80} 016112

\bibitem{Obizhaeva-Wang-2008-JFinM}
Obizhaeva A and Wang J 2010 {Optimal trading strategy and supply/demand
  dynamics} available at {SSRN}: http://ssrn.com/abstract=686168

\bibitem{Alfonsi-Fruth-Schied-2010-QF}
Alfonsi A, Fruth A and Schied A 2010 {\em Quant. Financ.\/} {\bf 10} 143--157

\bibitem{Zhou-2007-XXX}
Zhou W~X 2007 {Universal price impact functions of individual trades in an
  order-driven market} arXiv: 0708.3198v2

\bibitem{Gu-Chen-Zhou-2007-EPJB}
Gu G~F, Chen W and Zhou W~X 2007 {\em Eur. Phys. J. B\/} {\bf 57} 81--87

\bibitem{Jiang-Chen-Zhou-2008-PA}
Jiang Z~Q, Chen W and Zhou W~X 2008 {\em Physica A\/} {\bf 387} 5818--5825

\bibitem{Mu-Chen-Kertesz-Zhou-2009-EPJB}
Mu G~H, Chen W, Kert{\'e}sz J and Zhou W~X 2009 {\em Eur. Phys. J. B\/} {\bf
  68} 145--152

\bibitem{Bollen-Inder-2002-JEF}
Bollen B and Inder B 2002 {\em J. Emp. Financ.\/} {\bf 9} 551--562

\bibitem{Andersen-Bollerslev-Diebold-Ebens-2001-JFE}
Andersen T, Bollerslev T, Diebold F and Ebens H 2001 {\em J. Financ. Econ.\/}
  {\bf 61} 43--76

\bibitem{Andersen-Bollerslev-Diebold-Labys-2001-JASA}
Andersen T~G, Bollerslev T, Diebold F~X and Labys P 2001 {\em J. Am. Stat.
  Assoc.\/} {\bf 96} 42--55

\bibitem{Plerou-Gopikrishnan-Gabaix-Stanley-2002-PRE}
Plerou V, Gopikrishnan P, Gabaix X and Stanley H~E 2002 {\em Phys. Rev. E\/}
  {\bf 66} 027104

\bibitem{Zhou-Sornette-2004a-PA}
Zhou W~X and Sornette D 2004 {\em Physica A\/} {\bf 337} 243--268

\bibitem{Farmer-Gillemot-Lillo-Mike-Sen-2004-QF}
Farmer J~D, Gillemot L, Lillo F, Mike S and Sen A 2004 {\em Quant. Financ.\/}
  {\bf 4} 383--397

\bibitem{Biais-Hillion-Spatt-1995-JF}
Biais B, Hillion P and Spatt C 1995 {\em J. Financ.\/} {\bf 50} 1655--1689

\bibitem{Smith-Farmer-Gillemot-Krishnamurthy-2003-QF}
Smith E, Farmer J~D, Gillemot L and Krishnamurthy S 2003 {\em Quant. Financ.\/}
  {\bf 3} 481--514

\bibitem{Barber-Lee-Liu-Odean-2008-RFS}
Barber B~M, Lee Y, Liu Y and Odean T 2009 {\em Rev. Financ. Stud.\/} {\bf 22}
  609--632

\end{thebibliography}

\end{document}